\documentclass[12pt]{article}
\usepackage{amsmath}
\usepackage{amssymb}
\usepackage{amsthm}
\usepackage{amsfonts}
\usepackage{eurosym}
\usepackage{graphicx}
\usepackage{booktabs}
\usepackage{float}
\usepackage{subfig}
\usepackage{comment}
\usepackage{quoting}
\usepackage{multirow}
\usepackage{enumerate}
\usepackage{setspace}
\usepackage{pdflscape}
\usepackage[backend=biber, style=authoryear, maxcitenames=4, maxbibnames=99]{biblatex}
\usepackage{xcolor}
\usepackage{hyperref}
\usepackage{caption}
\usepackage{subcaption}
\usepackage{subfig}
\usepackage{titlesec}
\usepackage[table]{xcolor}
\usepackage{array}
\usepackage{enumitem}
\usepackage{threeparttable}
\usepackage{amsthm}

\titleformat{\section}
  {\normalfont\large\bfseries}
  {\thesection}{1em}{}

\usepackage{tikz}

\hypersetup{
    colorlinks=true,
    linkcolor=blue,
    filecolor=magenta, 
    citecolor=blue,
    urlcolor=blue,
    pdftitle={Overleaf Example},
    pdfpagemode=FullScreen,
}

\DeclareFieldFormat{labelyear}{\hyperlink{cite.\thefield{entrykey}}{\textcolor{blue}{#1}}}
\DeclareFieldFormat{shorthand}{\hyperlink{cite.\thefield{entrykey}}{\textcolor{blue}{#1}}}

\DeclareCiteCommand{\cite}
  {\usebibmacro{prenote}}
  {\usebibmacro{citeindex}%
   \textcolor{blue}{\hyperlink{cite.\thefield{entrykey}}{\printnames{labelname} (\printfield{year})}}}
  {\multicitedelim}
  {\usebibmacro{postnote}}

\AtEveryBibitem{%
  \stepcounter{citecount}\hypertarget{cite.\thefield{entrykey}}{}%
}

\newtheorem{lemma}{{\bf \sc Lemma}}
\newtheorem{corollary}{{\bf \sc Corollary}}
\newtheorem{proposition}{{\bf \sc Proposition}}

\newtheorem{definition}{{\bf \sc Definition}}

\usepackage[top=1in, bottom=1in, left=1in, right=1in]{geometry}

\begin{document}

\title{Strategic Expression, Popularity Traps, and Welfare in Social Media\footnote{\scriptsize First version: June, 2024 (available \href{https://papers.ssrn.com/sol3/papers.cfm?abstract_id=4881670}{here}).}}

\author{Zafer Kanik \and Zaruhi Hakobyan\thanks{
		\scriptsize{Kanik: Adam Smith Business School, University of Glasgow. Hakobyan: University of Luxembourg and CESifo. 
			Email: zafer.kanik@glasgow.ac.uk and zaruhi.hakobyan@uni.lu. We thank  Matthew Elliott, Matthew O. Jackson, and Alireza Tahbaz-Salehi for insightful discussions and their highly useful suggestions. We thank Francesco Squintani, Leeat Yariv, Sanjeev Goyal, Pau Milán, and participants of the Networks Workshop at the Barcelona Summer Forum 2024, University of Cambridge Economics Seminar Series, Dynamic Games and Applications Seminars 2025, IOS Economics Department Economics Seminars 2025, 14th ISDG Workshop 2025,  Armenian Economic Association Annual Meeting 2024 for helpful comments. Hakobyan acknowledges financial support from the FNR CORE C23/SC/18098569/POPULISM grant project.
            }}}

\date{\today}

\maketitle

\thispagestyle{empty}		
\setcounter{page}{0}

		\begin{abstract}

\begin{singlespace}
Social media platforms systematically reward popularity over authenticity, incentivizing users to strategically tailor their expression for attention. In this paper, we introduce (i) popularity as a strategic expression mechanism, distinct from the canonical mechanisms of conformity, learning, persuasion, and (mis)information transmission in social networks, and (ii) a utilitarian framework for measuring user
welfare that maps directly to observable platform metrics, filling a critical gap in the social media literature. In the model, agents hold fixed heterogeneous authentic opinions and derive utility gains from the popularity of their own posts---measured by \textit{likes} received, and utility gains (losses) from exposure to content that aligns with (diverges from) their authentic opinion. Social media interaction acts as a state-dependent welfare amplifier: light topics generate Pareto improvements, whereas intense topics make everyone worse off in a polarized society (e.g., political debates during elections). Moreover, strategic expression amplifies social media polarization during polarized events while dampening it during unified events (e.g., national celebrations). Consequently, strategic distortions magnify welfare outcomes, expanding aggregate gains in light topics while exacerbating losses in intense, polarized ones. Counterintuitively, strategic agents often face a \textit{popularity trap}: posting a more popular opinion is individually optimal, yet collective action by similar agents eliminates their authentic opinion from the platform, leaving them worse off than under the authentic-expression benchmark. Homophilic algorithms that match users with preferred content---widely used by platforms---discipline popularity-driven behavior, narrowing the popularity trap region and limiting its welfare effects.

    \end{singlespace}


		\end{abstract}

	\newpage

\section{Introduction}

\label{sec:introduction}

Social media platforms are defined by a distinctive feature: the quantification of social approval through ``likes." This mechanism is often tied to monetary or non-monetary rewards (\cite{eckles2016estimating}, \cite{burtch2022peer}, \cite{zeng2023impact}, \cite{aridor2024economics}, \cite{filippas2025production}), thereby incentivizing users to seek popularity---measured by
likes received---of their content.  When applied to opinion expression, this feature alters the strategic landscape by incentivizing agents to trade off popularity against authenticity. How does this strategic tension shape online polarization and user welfare from social media interaction?

We develop a model of strategic expression in social media and show that it generates redistributive and/or magnifying effects on social media polarization, individual utilities, and aggregate welfare. Moreover, homophilic algorithms widely used by platforms discipline rather than amplify these popularity-driven distortions, suggesting that targeted content delivery acts as a moderating force on strategic expression---and, contrary to common belief, may temper rather than fuel online polarization.

Our main contribution is twofold. First, we introduce an unexplored strategic expression channel in social networks, distinct from the canonical mechanisms of conformity, learning, belief updating, and persuasion (e.g., \cite{asch1955opinions, degroot1974reaching, bernheim1994theory,morris2001political,galeotti2009influencing, golub2012network,jadbabaie2012non,molavi2018theory, bloch2018rumors, pogorelskiy2019news, egorov2020persuasion,comola2024competing})\footnote{Readers are referred to \cite{pogorelskiy2019news} for a model of strategic information sharing in social media with learning in a voting framework.}. In our model, agents hold heterogeneous and fixed authentic opinions, and derive utility gains from the popularity of their own posts, and utility gains (losses) from exposure to content that aligns with (diverges from) their authentic opinion, featuring a trade-off between authenticity and popularity as well as externalities from others' content. This framework allows us to abstract away from the canonical mechanisms above and discover a new strategic landscape, shedding light on a disconnect between beliefs and online discourse. Relatedly, empirical studies by \cite{gentzkow2011ideological}, \cite{boxell2017greater}, and \cite{nyhan2023like} show that social media often has limited effects on beliefs. Besides, while a rich literature in behavioral economics documents how revealed preferences diverge from normative preferences---with \cite{beshears2008preferences} identifying the key factors driving this divergence---our model introduces a new such factor that specifically applies to online interaction in social networks: the pursuit of popularity. In this sense, our framework departs from existing models of content transmission online---whether through network diffusion (e.g., \cite{allcott2017social, vosoughi2018spread, steinert2015online}, \cite{abreu2019homophily}), algorithmic filtering (e.g., \cite{levy2021social, guess2023social, bakshy2015exposure}), or spread of misinformation (e.g., \cite{mostagir2022society, guriev2023curtailing, acemoglu2024model})---and instead models it as a strategic expression environment.

Second, we are the first to introduce a utilitarian framework for welfare analysis in social media that is defined directly over social media metrics: \textit{posts }and \textit{likes}. As emphasized in the review by \cite{aridor2024economics}, the existing literature evaluates welfare indirectly—through access choices, time use, or subjective well-being (e.g., \cite{allcott2020welfare,allcott2022digital,brynjolfsson2023digital,braghieri2024political})—rather than specifying utility directly over observable metrics. Our tractable utilitarian framework fills this critical gap in the social media literature, allowing us to measure individual utilities and aggregate welfare directly from social media interaction.\footnote{\cite{aridor2024economics} emphasize in their review: \begin{quotation}...While most existing work has concentrated on the time spent on social media, the content
consumed, and the nature of the engagement are also important. \end{quotation}

Existing work has primarily quantified welfare through stated-preference measures, such as willingness to pay or accept for data sharing and privacy, which face well-known challenges including uncertainty, externalities, and gaps between stated and revealed preferences. Our framework directly addresses this gap while isolating from learning/persuasion channels. Developing a hybrid microfoundation that
combines popularity-driven incentives with learning/persuasion is a natural direction for future research.}

In our analysis, we first focus on a representative social media environment in which each agent’s audience and exposure mirror the societal distribution of opinions. This benchmark abstracts from friendship and/or algorithmic biases—such as homophily—and clearly identifies the underlying mechanism. In Section \ref{sec:pvm}, we extend our analysis to algorithmic filtering by modeling a targeting rule widely used across social media platforms.  

In a multiple-opinion setting, first, we show that popularity-driven behavior does not generate new opinions on social media but rather reshuffles the existing ones: some views become overrepresented, while others become underrepresented, and some opinions may even vanish entirely from the platform. Crucially, the direction of these distortions is determined by the underlying distribution of opinions in society. 

Societal opinions are often sharply polarized, from intense political debates to light, taste-based topics, or sharply unified around broad consensus (e.g., national celebrations or responses to natural disasters). We apply our framework to these environments by introducing a tractable three-opinion space. In particular, we define an opinion space where each agent holds an authentic opinion in the set $\{-b, 0, b\}$, a polarized or neutral (central) view. The magnitude $|b|$ measures the topic intensity---representing the intensity of opposition between opinions---which serves as a key driver of strategic actions and resulting utility gains and losses.

We show a fundamental duality in how social media polarization differs from societal polarization: When society is already divided (e.g., political debates, immigration policy debates), popularity incentives amplify polarization in social media by inducing neutral agents to post opinionated content. Conversely, when society is highly unified (e.g., national celebrations, national sports events), the same incentives drive opinionated agents to converge toward neutral expression, generating amplified online unification. Social media thus acts as a non-linear magnifier, selectively distorting the perception of societal division depending on the nature of the event. Unlike existing work that studies online polarization (e.g., \cite{gentzkow2011ideological,bakshy2015exposure,barbera2015tweeting, campbell2019social, pogorelskiy2019news, callander2022cause,guess2023reshares,della2023affective,vosoughi2018spread,acemoglu2024model,bolletta2025dynamic}), our results focus on the role of popularity-driven strategic expression in it. Among these studies, \cite{campbell2019social} analyze polarization in a social media network with two news content types—mass-market and niche—where users differ in their preferred content to recommend. In their model, individuals recommend their preferred type of content when it is available in their network, and otherwise recommend the alternative one. They show that when the friendship network is not biased, content filtering tends to amplify mass-market content and reduce polarization at steady state--- defined as the difference between the mass and niche content recommendation probabilities. However, greater network connectivity and homophily may increase the prevalence of niche content and, in turn, increase polarization. In another related work, \cite{pogorelskiy2019news} analyzes content filtering in a similar two-type model within a voting framework. In their model, agents receive private signals and strategically share content based on their own partisan views and the composition of their audience on social media. Sharing decisions are strategic, whereas voting decisions are sincere and reflect Bayesian updating of beliefs about the state of the world. Different from these two models, our framework, in a multiple opinion setup, allows agents to strategically choose what to post (from a continuous opinion space) based on popularity and alignment incentives. Moreover, in a three-opinion setting, we show that the impact of strategic expression on polarization is state-dependent: polarization is amplified (dampened) when the underlying society is highly polarized (less polarized). 
 
To analyze the welfare (aggregate utility) implications, we evaluate individual utilities and welfare under the equilibrium outcome against two benchmarks: autarky (no social media activity on the given topic) and truthful expression (authentic posting by everyone).

In this framework with sticky opinions and popularity incentives, the welfare consequences are sharply state-dependent. For light topics, social media is welfare-improving because the utility gains from popularity and exposure to aligned content exceed the utility losses from exposure to misaligned yet low-stakes content. However, for intense topics, social media interaction is often harmful for some groups of agents; and for high-polarization events, intense topics create an environment in which everyone is worse off from social media interaction due to high misalignment costs. We also show that network density plays a purely welfare-amplifying role: denser networks increase exposure intensity, without altering equilibrium posting behavior. As a result, higher density magnifies welfare gains when social media interaction is welfare-improving, and amplifies welfare losses when it is welfare-reducing.

As emphasized by \cite{aridor2024economics}, an open question in the literature is estimating the effect of the intensity
of social media usage on welfare. Our framework provides a state-dependent explanation with non-monotonicity, showing how different exposure levels can generate either amplified welfare gains or amplified welfare losses depending on topic intensity and the degree of polarization on the given topic.

Crucially, in both high- and low-polarization cases, strategic posting behavior has heterogeneous effects for different opinion holders. Relatedly, we identify a phenomenon that we call the ``Popularity Trap'': for intense topics, social media activity can be utility-reducing for agents who strategically post a more popular opinion. Individually-optimal deviations by each such member cause their authentic opinion to disappear from social media, which amplifies their exposure to misaligned content, creating an equilibrium outcome where these strategic agents are worse off compared to the authentic expression benchmark. The popularity trap applies to neutral (opinionated) agents during high-(low-)polarization events. 

Our popularity trap complements the ``participation trap'' highlighted by \cite{bursztyn2025product}.
While their mechanism operates at the participation margin, ours arises at the margin of content creation within the platform. In their setting, agents feel compelled to join social media to avoid social exclusion, even if they derive negative utility from joining the platform that could be eliminated through coordinated exit. The popularity trap arising in our model could be avoided under coordinated authentic expression, suggesting some agents would be better off if they could collectively commit to authenticity in their posts.

Finally, our framework is distinct from the widely studied conformity models (e.g., \cite{asch1955opinions, bernheim1994theory,morris2001political,zhang2018fashion, buechel2015opinion, grabisch2019model, grabisch2020survey}). In conformity models, utility typically increases with similarity to others' actions. These incentives generate coordination benefits: an agent benefits when others also conform to the same norm. In our framework, the direction of externalities is reversed. Agents do not desire collective convergence; rather, they prefer to enjoy higher utility from popularity, but prefer that other agents who share their authentic opinion continue to express it authentically. This generates distinct utility  and welfare implications compared to conformity models, and also introduces the popularity trap.

The rest of the paper proceeds as follows. Section~\ref{model} introduces the model. Section~\ref{sec:expressed_opinions} characterizes the equilibrium posting behavior. Section~\ref{sec:polarization} introduces the three-opinion space and studies polarization, and Section~\ref{Polarization:Utility} fully characterizes the individual utilities and aggregate welfare in a setting isolated from network bias. Section \ref{sec:pvm} extends these results to a widely used, preference-based, platform algorithm setting. Section~\ref{sec:conclusion} concludes.

\section{The Model}
\label{model}
There is a finite set of agents \(\mathcal{N}=\{1,\dots,n\}\). On a given topic, each agent \(i\in\mathcal{N}\) holds an authentic opinion \(b_i\), which we take as exogenously given and fixed.\footnote{These fixed opinions represent individual-level sticky beliefs shaped by characteristics, experiences, identity, socioeconomic circumstances.} Each opinion $b_i$ lies in the opinion space $\mathcal{B}\equiv[-b,+b]\subset\mathbb{R}$. The magnitude $|b|>0$ denotes the topic intensity. Although the opinion space is continuous, the society exhibits a finite set of realized opinions $\mathcal{O}= \{b^{(1)},\dots,b^{(k)}\}\subseteq \mathcal{B}$, where $k<n$ holds, meaning that the number of realized opinions is less than the number of agents in society, so that for some realized opinions, there are multiple agents holding that same opinion.\footnote{The analysis remains similar if realized opinions are modeled as mutually exclusive (i.e., non-intersecting) intervals in \(\mathcal{B}=[-b,+b]\), rather than point values \(b^{(m)}\). As long as these intervals are non-overlapping and agents are assigned to them, the underlying mechanisms—and in particular the direction of endogenous reactions and the magnitude of agents having such endogenous reactions—in the model remain the same; but only the optimal deviation points within \(\mathcal{B}\) may differ, without generating additional insights.} For each realized opinion \(b^{(m)} \in \mathcal{O}\), \(G_{b^{(m)}}\) denotes the number of agents holding opinion \(b^{(m)}\), with
\(\sum_{b^{(m)} \in \mathcal{O}} G_{b^{(m)}} = n\).
We denote the corresponding set of agents by \(\mathcal{G}_{b^{(m)}}\). The opinion profile is the vector of agents' opinions \(\mathbf{b}=(b_1,\dots,b_n)'\).\footnote{Subscripts index agents (e.g., $b_i$), whereas superscripts index realized opinion types in $\mathcal{O}$ (e.g., $b^{(m)}$).} The \emph{society} is denoted by the pair \((\mathbf{b},\mathcal{N})\).

\paragraph{Social media behavior.}
The equilibrium concept is \emph{Subgame Perfect Nash Equilibrium} (SPNE). The sequence of events is as follows. 

At time \(t=0\), each agent simultaneously\footnote{An equivalent formulation is to define the timing and actions fully sequentially, as clarified in Footnote~\ref{footnote5}.} creates a social media post representing a view $c_i \in [-b,+b]\subset\mathbb{R}$. The vector of posts $\mathbf{c}=(c_1,\dots,c_n)'$ is the equilibrium expressed opinions on social media. The \emph{social media} is denoted by the pair $(\mathbf{c}, \mathcal{N})$.

At time \(t=1\), the post of each agent \(i\) is visible to the friendship (or follower) set of agent \(i\), denoted by \(\mathcal{A}_i \subset \mathcal{N}\setminus\{i\}\). This set consists of a positive $A_{i,b^{(m)}}$ number of agents holding opinion $b^{(m)}$ for each realized opinion $b^{(m)} \in \mathcal{O}$, where $0 < A_{i,b^{(m)}} = a_{i,b^{(m)}} G_{b^{(m)}} < G_{b^{(m)}}$. The total number of followers of agent $i$ is denoted by $A_i$, where $0 < A_i =|\mathcal{A}_i| = \sum_{b^{(m)}\in\mathcal{O}} A_{i,b^{(m)}} = \sum_{b^{(m)}\in\mathcal{O}} a_{i,b^{(m)}} G_{b^{(m)}}$. Each individual-opinion-specific parameter \(0 \leq a_{i,b^{(m)}} < 1\) is exogenous, and captures agent $i$’s audience reach among agents holding opinion $b^{(m)} \in \mathcal{O}$. The social media network is directed, implying that exposures between agents need not be reciprocal.

At time \(t=1\), posts are shown sequentially. Specifically, the reaction stage consists of \(n\) substeps: at each substep, exactly one agent is randomly selected (without replacement), and that agent’s post is shown to their followers. Each agent \(j \in \mathcal{A}_i\) viewing agent \(i\)’s post simultaneously decides whether to \emph{like} the post or not to react.\footnote{\label{footnote5}The timing of events at \(t=0\) could alternatively be modeled in \(n\) discrete steps. More precisely, at each substep, exactly one agent is randomly selected (without replacement) to create a post; denote this agent by \(i(t)\). After \(c_{i(t)}\) is posted, all agents in \(\mathcal{A}_{i(t)}\) who observe the post (i.e., the followers of \(i(t)\)) simultaneously (or similarly, sequentially) decide whether to like it or not to react. The outcome of this sequential formulation is equivalent to that of the simultaneous-posting and simultaneous-reactions we use. The equivalence follows from the fact that equilibrium posting and reaction behavior is independent of the ordering of agents’ actions, provided that follower sets remain fixed throughout the posting stage.}

The popularity of agent \(i\)'s post is measured by the total number of \textit{likes} it receives--- capturing social media appreciation, recognition, or influence:
    \[
    R_i= \sum_{j \in \mathcal{A}_i} r_{ji},
    \]
    where \(r_{ji}=1\) if agent \(j \in \mathcal{A}_i\) likes agent \(i\)’s post, and \(r_{ji}=0\) otherwise. $\mathbf{R} = [\, r_{ij} \,]_{i,j \in \mathcal{N}}$ denote the $n \times n$ matrix collecting all individual reactions. The action profile \((\mathbf{c}, \mathbf{R})\) constitutes an SPNE if no agent has a profitable deviation in any subgame.

\paragraph{Information structure.} Each agent \(i\) observes their own authentic opinion \(b_i\) and her total number of followers $A_{i,b^{(m)}}$ for each opinion group \(b^{(m)} \in \mathcal{O}\). This information constitutes the agent’s entire information set for strategic decision-making in the model. In particular, agents do not observe (or equivalently, do not necessarily observe) the exact members in their follower set \(\mathcal{A}_i\), and the opinions, audience compositions, or exact follower sets of any other agent \(j \in \mathcal{N} \setminus \{i\}\). This information structure captures a platform environment in which users have only an understanding of their own audience composition, while they might be lacking knowledge about the individuals, their views, or reach of others on social media.

Finally, it is common knowledge to everyone that all agents assign strictly positive and finite weight to each component of the utility function introduced below. Each agent observes their own utility weights exactly, but does not (necessarily) observe the exact utility weights of other agents.

\paragraph{Utility function.}

Both post creation and likes are determined according to the utility function below with positive and finite weight parameters $\omega^p_i \in (0,\infty)$,  $\omega^d_i \in (0,\infty)$, and $\omega^a_i \in (0,\infty)$. At any equilibrium $(\mathbf{c}, \mathbf{R})$,

\begin{equation}
\label{utilityf}
    U_i(\mathbf{c},\mathbf{R}) 
    = H_i + \omega^p_i R_i     + \omega^a_i \!\!\sum_{\substack{j \in \mathcal{N}_i \\ c_j = b_i}} \! R_j 
    - \omega^d_i \!\!\sum_{\substack{j \in \mathcal{N}_i \\ c_j \neq b_i}} R_j |b_i - c_j|
 .
\end{equation}

$H_i$ denotes agent $i$'s baseline utility, capturing all payoffs except those generated by social media activities on the specific ``topic" modeled. 

The term $\omega^p_i R_i$  captures popularity gains from her own post. A higher \(\omega^p_i\) corresponds to higher utility gains from each additional like she receives.

The remaining components capture alignment-(misalignment-)based utility gains (losses) from content exposure. Agent \(i\)’s exposure set is defined as $\mathcal{N}_i
\;\equiv\;
\{i\}
\;\cup\;
\{\, j \in \mathcal{N} \setminus \{i\} \;:\; i \in \mathcal{A}_j \,\}$. That is, \(\mathcal{N}_i\) consists of agent \(i\) herself (exposed to her own post) together with all agents whose posts are visible to \(i\) on the platform. 

The term $\omega^a_i \sum_{j \in \mathcal{N}_i, c_j = b_i} R_j$ captures the utility gains agent $i$ derives from exposure to content that aligns with her authentic opinion $b_i$. Crucially, this benefit is scaled by $R_j$, implying that agents value not just the presence of aligned views in their feed, but their resonance in the platform. In this specification, the popularity of an aligned post serves as a proxy for social proof or societal validation: seeing a post that agrees with one's view go viral generates significantly higher utility than seeing the same opinion ignored, as the former signals widespread societal approval of the agent's beliefs. For example, a user holding a specific political stance derives satisfaction not merely from reading a supportive argument, but from observing that thousands of others have publicly endorsed it with "likes," reinforcing her sense of belonging and correctness. The parameter $\omega^a$ governs the intensity of this validation motive.

The fourth component, $\omega^d_i \sum_{j \in \mathcal{N}_i} R_j |b_i - c_j|$, captures the disutility agent $i$ experiences from exposure to posts that diverge from her authentic opinion. This cost is amplified by $R_j$. A post \(c_j\) that has both a greater distance from agent \(i\)’s opinion and is widely endorsed generates a larger utility loss, because it signals not merely a difference of opinion, but strong opposition to the agent’s authentic belief.
We allow $\omega^d_i$ and $\omega^a_i$ to differ, acknowledging that agents may exhibit asymmetric sensitivities to alignment versus misalignment.

\section{Expressed Opinions and Distorted Utilities}
\label{sec:expressed_opinions}

In this environment, a post \(c_i \in [-b,+b]\) that differs from \(b_i\) reflects inauthentic, popularity-driven posting. Although identifying whether an expressed opinion is authentic or inauthentic is difficult in reality, our utility framework offers a way to conceptually separate the two, as discussed after our first result.

\begin{lemma}
\label{prop:SPNE}
In a given society $(\mathbf{b},\mathcal{N})$ at equilibrium $(\mathbf{c},\mathbf{R})$, each agent's post \(c_i\) corresponds to one of the realized opinions in society, i.e.,

\[
c_i \in \mathcal{O} = \{b^{(1)}, \dots, b^{(k)}\}, 
\qquad \forall i \in \mathcal{N}.
\]

\end{lemma}

Recall that the strategy space allows agents to post any opinion \( c_i \in [-b, +b] \). However, Lemma~\ref{prop:SPNE} implies that popularity incentives do not generate novel viewpoints: no equilibrium post lies strictly between the discrete elements of the realized opinion set \( \mathcal{O} \). Thus, despite the continuous choice set, expressed opinions are always constrained to the existing finite set of authentic beliefs. Popularity-driven social media, therefore, shapes outcomes not by creating new opinions, but by distorting the frequencies of existing ones.

Lemma~\ref{prop:SPNE} relies on a duality in equilibrium: agents’ authentic preferences are reflected in their \textit{likes} on others' posts but not necessarily in their own posts. At equilibrium, agent \(i\) likes agent \(j\)'s post if and only if the post aligns with her own opinion, \(c_j = b_i\). For any \(c_j \neq b_i\), liking \(j\)'s post yields disutility \(-\omega^d_i |b_i - c_j|\), and therefore $i$ chooses not to react ($r_{ij}=0$). In contrast, when \(j\)'s post aligns with \(i\)'s opinion (\(c_j = b_i\)), liking provides a positive payoff \(\omega^a_i\). Hence, \(r_{ij}=1\) if and only if \(c_j=b_i\).

However, an agent \(i\) may choose to post \(c_i \neq b_i\) in order to attract more attention. Recall that an agent $i$'s post's visibility to opinion group $b^{(m)}\in\mathcal{O}$ is equal to $A_{i,b^{(m)}}=a_{i,b^{(m)}}G_{b^{(m)}}$. Then, based on the opinion-match liking behavior explained above, $R_i=A_{i,b^{(m)}}$ holds for any given $c_i=b^{(m)}$. 

The baseline utility, \(H_i\), and the utility component \(F_i\)---that captures agent \(i\)’s social media utility on the given topic except those derived from her own post--- are both independent from the agent's own posting decision, and therefore do not play a role in the posting decision of agent $i$. As a result, agent \(i\)'s equilibrium post \(c_i\) solves:
\[
\max_{c_i \in [-b,+b]} A_{i,c_i}
\bigl(\omega^p_i + \omega^a_i \mathbf{1}_{\{c_i = b_i\}} - \omega^d_i |b_i - c_i|\bigr).
\]

By revealing the authentic opinion (\(c_i = b_i\)), $i$ receives $U_i= H_i + F_i + A_{i,b_i} \cdot (\omega^p_i + \omega^a_i)$. Alternatively, \(i\) may post (\(c_i = b^{(m)} \neq b_i\)) and earn $U_i=H_i + F_i + A_{i,b^{(m)}} \cdot (\omega^p_i - \omega^d_i |b_i - b^{(m)} |)$. Moreover, posting an opinion $ c_i \notin \mathcal{O}$ that receives no likes is always inferior to posting an authentic content, and consequently, agent $i$ posts authentic if and only if
\[
(\omega^p_i+\omega^a_i)A_{i,b_i} 
\ge 
\max_{b^{(m)} \in \mathcal{O} \setminus \{b_i\}} 
A_{i,b^{(m)}}\big(\omega^p_i-\omega^d_i|b_i-b^{(m)}|\big).
\]

Otherwise, agent $i$ posts the alternative opinion $b^{(m)} \neq b_i$ in the set $\mathcal{O}$ and maximizes $A_{i,b^{(m)}}\big(\omega^p_i-\omega^d_i|b_i-b^{(m)}|\big)$. As a result, strategic posting, by reshuffling opinion  distribution, directly affects individual utilities and welfare on social media through self-posts and exposure.

While the distortion mechanism applies to various distributions of societal opinions, in practice, societies often exhibit specific opinion distributions. For instance, during national celebrations or collective tragedies, a neutral opinion is held by a large majority, reflecting a highly unified society. In others, divisive events—ranging from high-stakes debates to low-stake daily-life or taste-based discussionson on food, art, fashion,...---are characterized by polarization, with society mainly split across opposing positions. To capture these canonical environments and to characterize who benefits and who loses from social media interaction, we specialize to a three-opinion space in the remainder of the paper.

\section{Polarization and Welfare on Social Media}
\label{sec:polarization}

The set of realized opinions in society is: $\mathcal{O}=\{-b,\,0,\,+b\}$. The opinions \(-b\) and \(+b\) represent two opposing polarized views, while \(0\) denotes a neutral position. Let \(G_-\), \(G_0\), and \(G_+\) denote the population sizes of agents holding opinions \(-b\), \(0\), and \(+b\), respectively, with $G_- + G_0 + G_+ = n$; and let \(\mathcal{G}_-\), \(\mathcal{G}_0\), and \(\mathcal{G}_+\) denote the respective sets. For tractability, we focus on a symmetric configuration:  $G_- = G_+ = \frac{n - G_0}{2}$. The neutral is the largest (smallest) opinion group for \(G_0 > n/3\) (\(G_0 < n/3\)), in which case we write \(G_0 := G_{\max}\) (\(G_0 := G_{\min}\)).\footnote{The \href{https://papers.ssrn.com/sol3/papers.cfm?abstract_id=4881670}{working paper} version
  provides a result in a more general setting in which the three group sizes are strictly ordered as \(G_{\max} > G_{\text{med}} > G_{\min} > 0\). Strategic posting is less prevalent when the neutral group is of intermediate size ($G_0=G_{\text{med}}$). Accordingly, $(G_0=G_{\min})$ or $(G_0=G_{\max})$ are the most informative for studying popularity-driven behavior.}

The equilibrium group sizes representing opinions \(-b\), \(0\), and \(+b\) on social media are denoted by \(C_-\), \(C_0\), and \(C_+\), respectively. Following Lemma~\ref{prop:SPNE}, \(C_- + C_0 + C_+ = n\) holds. To avoid divisibility issues, we assume that \(n\) and \(G_0\) are even numbers. Moreover, if there exists some neutral agents who are indifferent between posting one of the opinionated views \(-b\) and \(+b\), we assume that the number of such indifferent neutral agents is always even,  and the equilibrium posts of these agents are evenly split between \(C_-\) and \(C_+\).

\begin{definition}
\label{def1}
For fixed $n$ and \(|b|\), social media \((\mathbf{c},\mathcal{N})\) is weakly more (less) polarized than society \((\mathbf{b},\mathcal{N})\) if and only if \(C_0 \leq G_0\) (\(C_0 \geq G_0\))
and strictly more (less) polarized if and only if \(C_0 < G_0\) (\(C_0 > G_0\)). 
\end{definition}

To isolate the distortion channel from network-bias (or algorithmic-bias), we first focus on the representative social media environment defined below. Section \ref{sec:pvm} extends results towards algorithmic-bias.\footnote{The \href{https://papers.ssrn.com/sol3/papers.cfm?abstract_id=4881670}{working paper} version  includes other results on how strategic expression can be driven by some additional network properties including network positions of agents (e.g., influencers vs. periphery).} 

Social media is \emph{representative} if each agent’s both follower composition and her exposure set—excluding herself—reflect the distribution of societal opinions.\footnote{This representative structure can alternatively be interpreted in terms of agents’ probabilistic beliefs about the \emph{composition} of their follower set. When users do not know the exact group-size composition of their followers, a natural assumption is that they assign probabilities over the shares of different opinion groups according to the population distribution of opinions. Combined with opinion-alignment–driven deterministic liking behavior, this probabilistic interpretation yields results with the same qualitative insights as our deterministic formulation.} Formally:
\[
0 < a_{i,b^{(m)}} = a_i < 1 \quad \text{and} \quad 
\frac{|\{j \in \mathcal{N}_i \setminus \{i\} : b_j = b^{(m)}\}|}{|\mathcal{N}_i \setminus \{i\}|}
=
\frac{G_{b^{(m)}}}{n}
\quad \forall b^{(m)} \in \mathcal{O}, \forall i \in \mathcal{N}.\] 

Individual-specific parameters $a_i$ and $|\mathcal{N}_i|$ are allowed to differ across agents. Proposition~\ref{prop1} shows \textit{event-specific} polarization effects.

\begin{proposition}
	\label{prop1}	
	 For fixed $n$ and $|b|$:

\begin{itemize}

 \item[(i)] for $G_0 := G_{\min}$ (a high-polarization event), social media is weakly more polarized than society, and strictly more polarized if and only if there exists neutral agents with sufficiently strong popularity incentives satisfying: 
$$\omega^p_i > (\omega^p_i)^*=\frac{2 \omega^a_i G_0 + \omega^d_i |b| (n - G_0)}{n - 3 G_0}, \text{or equivalently } \quad 
G_0 < G_0^\ast
=
\frac{(\omega_i^p-\omega_i^d|b|)\,n}{3\omega_i^p-\omega_i^d|b|+2\omega_i^a},$$

    \item[(ii)] for $G_0 := G_{\max}$ (a low-polarization event), social media is weakly less polarized than society, and strictly less polarized if and only if there exists polarized agents---whose authentic opinion is either $-b$ or $+b$---with sufficiently strong popularity incentives satisfying: 
    $$\omega^p_i > \frac{\omega^a_i (n - G_0) + 2 \omega^d_i |b| G_0}{3 G_0 - n}, \text{ or equivalently} \quad G_0>G_0^{\ast\ast}
=
\frac{(\omega_i^p+\omega_i^a)\,n}{3\omega_i^p-2\omega_i^d|b|+\omega_i^a}.$$
  \end{itemize}

\end{proposition}

Proposition~\ref{prop1} shows that social media amplifies polarization during a high-polarized event and
dampens polarization during a low-polarized event.
When neutral agents constitute a sufficiently large group (\(G_0 > G_0^{\ast\ast}\)), some opinionated agents strategically create neutral posts, making social media appear more unified than the society, e.g., through widely shared posts promoting solidarity or collective emotion, leading to amplified endogenous echo chambers around neutral themes. Conversely, when neutrals constitute a small group (\(G_0 < G_0^{\ast} \)), some neutral agents having sufficiently large popularity incentives strategically post opinionated, rendering social media more polarized than society.

\begin{figure}[h!]
    \centering
    \includegraphics[width=0.7\linewidth]{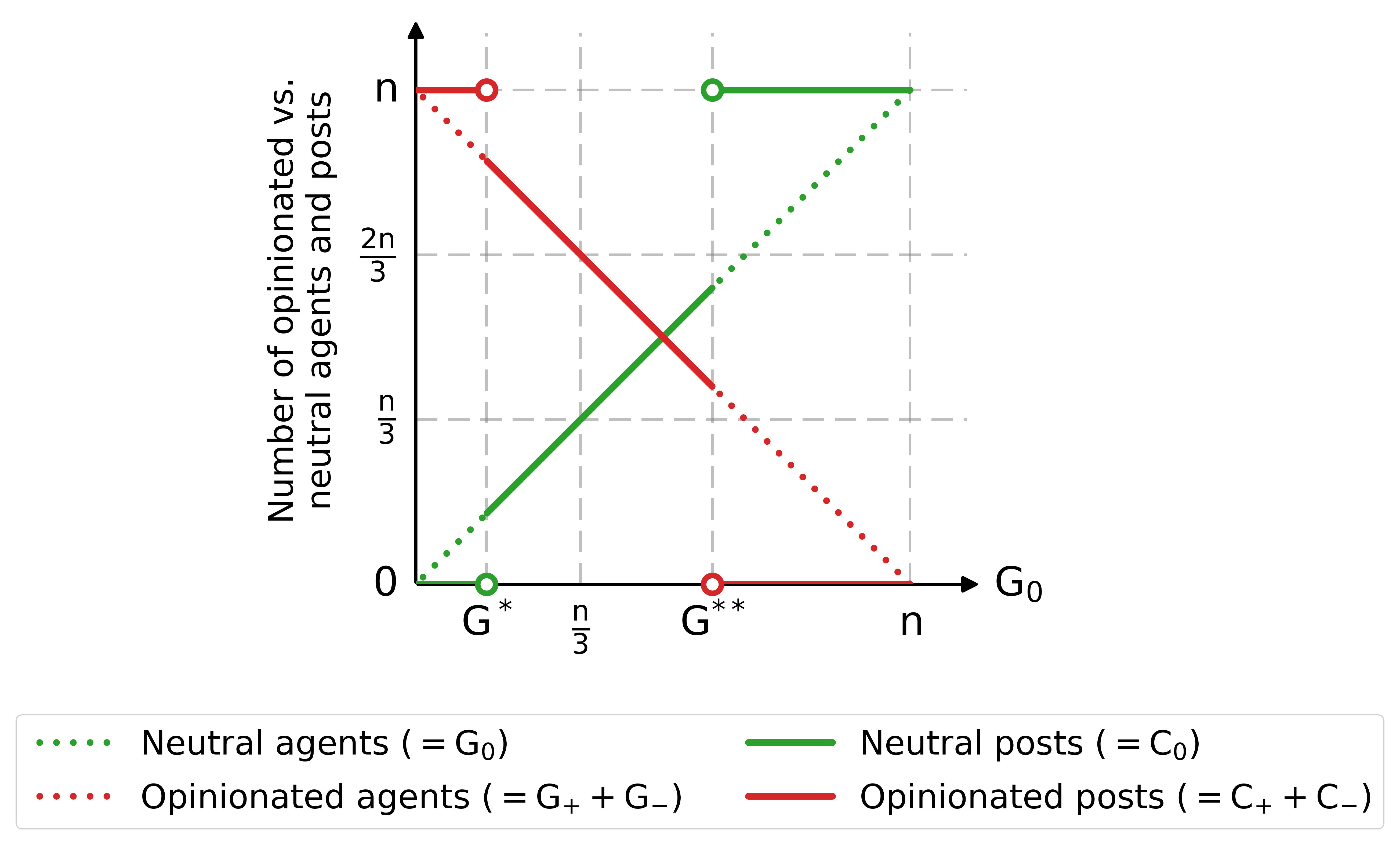}
    \caption{Authentic Opinions versus Expressed Opinions.}
\label{fig:popdriven}
    \vspace{0.5em}
   \raggedright{  {\footnotesize
    \raggedright
    \emph{Notes:} For the parameterization \(\omega^p=2\) and \(\omega^a=\omega^d=1\) (i.e., agents value one additional like on their own post twice as much as the intrinsic utility/disutility from alignment/misalignment), $|b|=1$, and \(n=100\), the implied thresholds are \(G_0^{\ast} \approx 14\) and \(G_0^{\ast\ast} = 60\). The x-axis plots $G_0 \in \{0, 2, 4, \dots, n\}$; where $G_0$ and $n$ are assumed to be even and deviating agents are assumed to be splitted evenly.}}
\end{figure}

Figure \ref{fig:popdriven} illustrates such event-specific differences under homogeneous utility parameters ($\omega^p_i =\omega^p$, $\omega^a_i = \omega^a$ and $\omega^d_i =\omega^d \quad \forall i \in \mathcal{N}$). The dashed green and red lines represent, respectively, the number of neutral agents ($G_0$) and the total number of polarized agents ($n - G_0 = G_+ + G_-$) in society, and the solid green and red lines depict neutral vs. polarized posts on social media. We now present a corollary under homogeneous utility parameters, as in Figure~\ref{fig:popdriven}.

\begin{corollary}
\label{unification}

For $G_0:=G_{\min}$, let $G^{\text{neutral}}$ denote the threshold size of the neutral group $G_0$ such that if \(G_0 < G^{\text{neutral}}\), all neutral agents create opinionated posts on social media. For $G_0:=G_{\max}$, let $G^{\text{opin}}$ denote the threshold size of the opinionated group ($=G_-+G_+$) such that for $(G_-+G_+) < G^{\text{opin}}$, all opinionated agents create neutral posts on social media. Then, for fixed \(n\), \(|b|\), \(\omega^p\), \(\omega^a\), and \(\omega^d\), it holds that:
\[
G^{\text{neutral}} < G^{\text{opin}}.
\]
\end{corollary}

Corollary \ref{unification} shows that amplified unification on social media during unified events (i.e., sufficiently large \(G_0\)) emerges more easily than amplified polarization during polarized events (i.e., sufficiently small \(G_0\)). This follows from the fact that popularity-driven posting is more prevalent in unifying events than in polarizing ones.

\section{Individual Utilities and Welfare}

\label{Polarization:Utility}

Throughout this section, the information setting remains as same as in Section \ref{model}, and we impose homogeneity:
\[
H_i=\overline{H},\qquad
\omega_i^p=\omega^p,\quad \omega_i^a=\omega^a,\quad \omega_i^d=\omega^d,\qquad
a_i=a, \qquad |\mathcal N_i|=an+1
\qquad \forall i\in\mathcal{N}.
\]

Our analysis uncovers a fundamental property of digital platforms: even in the absence of strategic distortions, the welfare implications of opinion expression are structurally ambiguous. Therefore, first, we present a simulation result comparing social media utility under authentic expression where every agent truthfully posts their opinion ($c_i = b_i$ for all $i \in \mathcal{N}$) with the baseline utility of autarky (no social media activity for the given topic) with $U_i^{\mathrm{aut}} = \overline{H}$.

\begin{figure}[t!]
    \centering
    \includegraphics[width=1.0\textwidth]{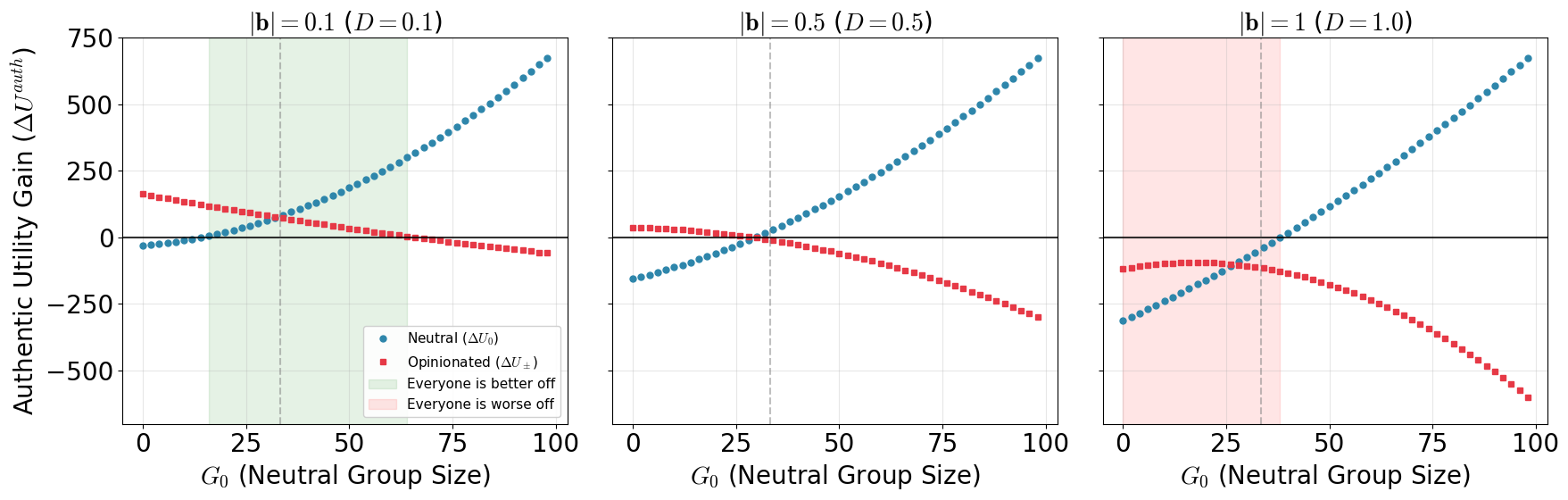}
    \caption{\footnotesize \textbf{Authentic Expression by All Agents vs. Autarky.} 
    The figure plots the utility gain for each agent type under authentic expression ($c_j = b_j \forall j \in \mathcal{N}$) by all agents ($\Delta U^{\mathrm{auth}}_i$), for a neutral agent (blue line) and opinionated agent (red line).
    \textit{Left Panel ($|b|=0.1$):} For low-intensity topics, authentic participation yields positive utility (a Pareto improvement over autarky) for all agents regardless of group size.
    \textit{Right Panel ($|b|=1.0$):} For high-intensity topics, minority groups suffer negative utility even under authentic expression, as the cost of exposure to opposing views outweighs popularity and alignment benefits.}
    \label{fig:authentic_vs_autarky}
\end{figure}

For high-intensity topics (Right Panel of Figure \ref{fig:authentic_vs_autarky}, $|b|=1.0$), in high-polarization events (low $G_0$), the overwhelming volume of opposing content drives a welfare loss for all agents (red region): Every agent is strictly worse off than under autarky. As the neutral group expands, neutral agents eventually gain, but at the direct expense of opinionated agents. Thus, high-intensity topics generate collective harm or asymmetric winners, instead of a Pareto improvement.

For low-intensity topics (Left Panel of Figure \ref{fig:authentic_vs_autarky}, $|b|=0.1$), a collective gain (green region) becomes attainable. However, this requires specific structural conditions. In high-polarization events (low $G_0$), neutral agents suffer net losses as a minority. Conversely, in low-polarization events (high $G_0$), opinionated agents lose out. Pareto improvement therefore emerges only in intermediate polarization events: sufficiently diverse population that avoid excessive misalignment costs for each type. 

Altogether, the simulation shows that in polarized or unified events, there are often winners and losers from social media interaction.

\begin{figure}[t!]
    \centering   \includegraphics[width=1.0\textwidth]{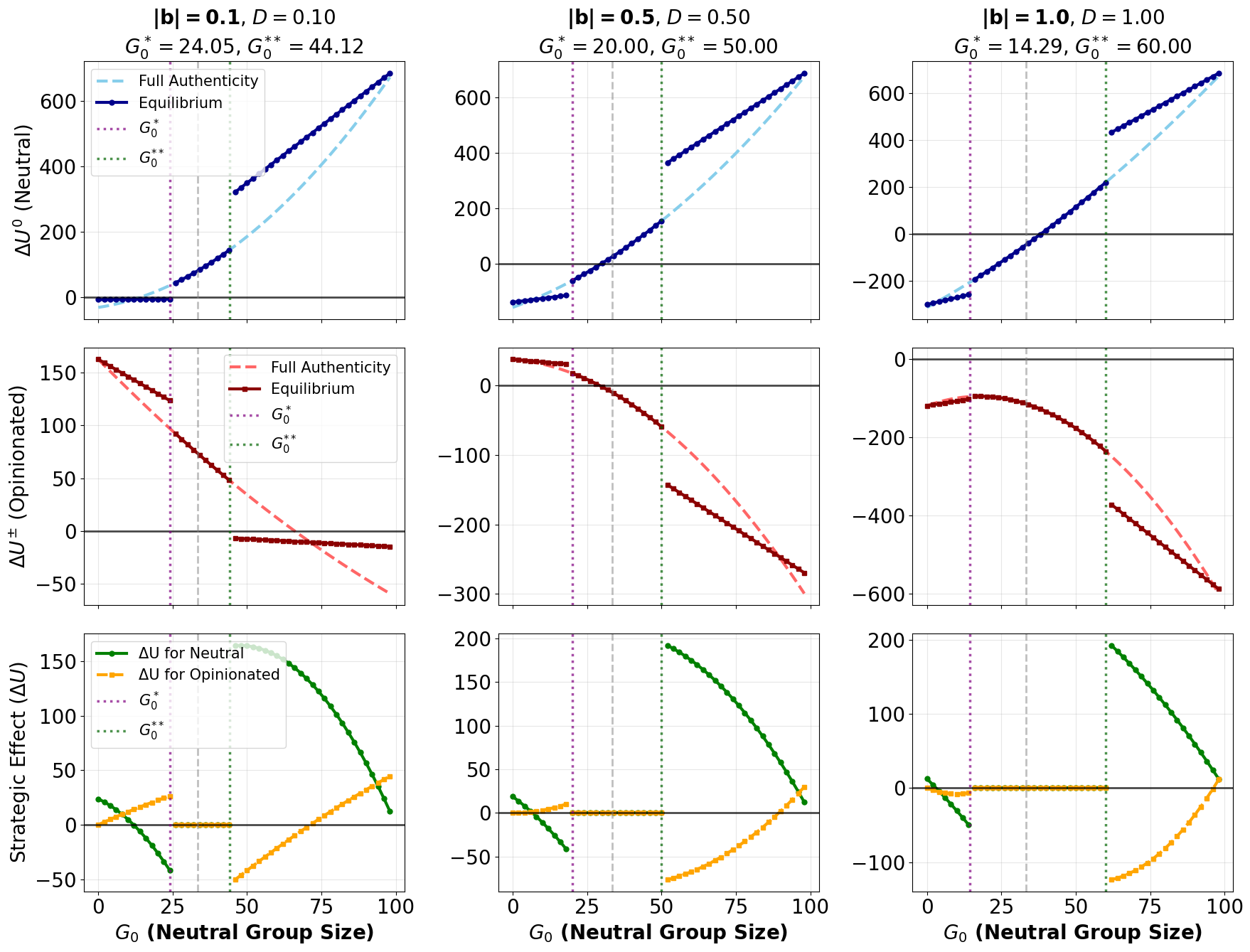}
    \caption{\footnotesize \textbf{Equilibrium Outcomes vs. Authenticity Benchmark.} 
    The figure decomposes welfare effects into Equilibrium levels (dark lines) and Authentic benchmark levels (light lines).
    \textbf{Rows 1 \& 2:} Utility levels under each case for neutral and opinionated agents, respectively.
    \textbf{Row 3:} The \textit{Strategic Effect} ($\widehat{\Delta} = U^{eq} - U^{auth}$), isolating the gains/losses under strategic actions. In High-Polarization regimes (left of each plot), neutral agents' posting opinionated content often cause a utility loss for neutrals (Green line $<0$). In Low-Polarization regimes (right of each plot), opinionated agents' posting neutral content benefits neutrals at the expense of opinionated agents (Green $>0$, Orange $<0$).
    Parameters: $n=100$, varying intensity $|b| \in \{0.1, 0.5, 1.0\}$.}
    \label{fig:eq_vs_full_authentic}
\end{figure}

Having established the baseline welfare properties, we now isolate the specific impact of strategic behavior. Figure \ref{fig:eq_vs_full_authentic} compares the equilibrium outcomes against the counterfactual of authenticity, where every agent is assumed to post truthfully.

Simulation findings in Figure \ref{fig:eq_vs_full_authentic}  reveal that strategic incentives systematically redistribute welfare across groups, and also generating ``popularity traps" for different groups of agents depending on societal opinion distributions. The third row explicitly plots the strategic effect ($\widehat{\Delta}_i = U^{eq}_i - U^{auth}_i$), capturing the net utility gain or loss generated by strategic actions.

\textit{High-Polarization Events (Left Side of Every Plot).} 
In the regime where neutral agents are the minimum group size ($G_0 < n/3$) and sufficiently small, the relative returns to popularity induce a strategic deviation toward polarized opinions. As shown in the third row (Green line, Column 1), this equilibrium outcome frequently yields a net utility loss for neutral agents ($\widehat{\Delta}^0 < 0$). The collective elimination of neutral content exposes these agents to higher misalignment costs relative to the truthful benchmark, and eventually lead to lower utility levels for these agents compared to the authentic benchmark case. This characterizes the popularity trap for neutrals.

\textit{Low-Polarization Events (Right Side of Every Plot).} 
In the regime where neutral agents dominate ($G_0 > n/3$) with a sufficiently high size, opinionated agents strategically moderate to appeal to the center. This behavior generates a stark asymmetry in welfare effects. Neutral agents (Green line) experience a substantial utility rise ($\widehat{\Delta}^0 > 0$) as the platform is flooded with aligned content from all users. Conversely, opinionated agents (Orange line) often suffer a popularity trap ($\widehat{\Delta}^\pm < 0$).

We now systematically compare the utility difference between the popularity-driven equilibrium and the authentic benchmark. Before showing formal existence-based results, first we provide the full characterization of potential regimes.

\paragraph{Popularity-driven equilibrium. } In the high-polarization event, the \textit{PE} (henceforth called as \textbf{PE}) refers to the unique outcome where neutral agents strategically post opinionated content, while opinionated agents continue to post authentically. Conversely, in the low-polarization event, PE refers to the unique equilibrium outcome where opinionated agents strategically post neutral content, while neutral agents continue to post authentically.

Let $\omega^d|b|=D \in(0,\infty)$ denote the unit misalignment (or distance) cost. The comparisons below are primarily governed by the magnitude of $D$. The net utility gains at the PE, if reached as an outcome, relative to the gains in the authenticity benchmark for an agent of type $k \in \{0, \pm\}$ be defined as $\Delta U^k := U^{\mathrm{eq},k} - U^{\mathrm{auth},k}$. 

\subsection{High-Polarization Event}

First, recall that in a high-polarization event, neutral agents post opinionated content if and only if the misalignment cost is sufficiently low. We denote this threshold as $D^*$:
\[
D < D^\ast:=\frac{\omega^p(n-3G_0)-2\omega^aG_0}{n-G_0}.
\]
If $D \ge D^*$, agents post authentically, and utilities coincide with the benchmark. When $D < D^*$, PE exists and unique, and in this case two additional thresholds arise:

\begin{itemize}

    \item \textit{The neutral benefit threshold. $D_0^{\mathrm{high}}$} marks the maximum misalignment cost under which neutral agents remain better off in the PE compared to the authenticity benchmark. If $D < D_0^{\mathrm{high}}$, the popularity gains from expanded reach outweigh the cost of exposure to a more intensely polarized feed. If $D > D_0^{\mathrm{high}}$ (but remains below $D^*$), they suffer a net loss despite choosing to post an opinionated content.

    \medskip
    
    \item \textit{The opinionated benefit threshold. $D_\pm^{\mathrm{high}}$} marks the maximum misalignment cost under which opinionated agents are better off ($\Delta U^\pm>0$). If $D < D_\pm^{\mathrm{high}}$, the benefit of being exposed to more aligned content outweighs the loss of the neutral buffer relative to fully opposing content (because a half of neutrals post the opposite view in the PE). If $D > D_\pm^{\mathrm{high}}$, the loss of the neutral buffer dominates, leaving them worse off.
\end{itemize}

There exists four distinct potential regimes:

\medskip
\textit{(a) A potential region in which everyone is better off.}
When the misalignment cost is sufficiently low, the strategic posting equilibrium makes every agent better off. In this region, the gains from expanded likes (for neutrals) and increased aligned content (for opinionated agents) overwhelm the relatively low distance costs in a more polarized environment. Formally,
\[
\Delta U^0 > 0 \text{ and } \Delta U^\pm > 0 
\;\Longleftrightarrow\;
0 \le D < \min\!\left\{ D^*,\; D_\pm^{\text{high}},\; D_0^{\text{high}} \right\}.
\]

\medskip

\textit{(b) A potential region in which only the neutral agents are worse off.}
For intermediate misalignment costs, PE harms neutral agents while benefiting opinionated ones. Opinionated-posting remains individually rational for neutrals ($D < D^*$), but the cost of the resulting sufficiently intense polarized feed is too high relative to the popularity gain. Opinionated agents, however, still enjoy the net benefit of the influx of aligned posts. This occurs precisely when
\[
D^{\text{high}}_0 < D < \min\!\left\{ D^{\text{high}}_\pm,\; D^* \right\}.
\]

\medskip

\textit{(c) A potential region in which only the opinionated agents are worse off.}

This could happen if the misalignment cost is low enough to benefit strategic neutrals (who gain sufficient popularity) but high enough to harm opinionated agents. This outcome holds if and only if
\[
D^{\text{high}}_\pm < D < \min\!\left\{ D_0^{\text{high}},\; D^* \right\}.
\]

\medskip

\textit{(d) A potential region in which all agents are worse off.}
Here, the misalignment cost is low enough to induce strategic posting ($D < D^*$), but high enough that it overwhelms the benefits for everyone. This collective loss arises if and only if
\[
\max\!\left\{ D^{\text{high}}_\pm,\; D_0^{\text{high}} \right\} < D < D^*.
\]

\medskip

Outside the PE region---that is, when $D \ge D^*$---agents post authentically, and utilities coincide under both benchmarks. 

Among these distinct outcomes, regimes (b) and (d) constitute a \textit{popularity trap} for neutral agents. The following result formalizes the existence of this region.

\begin{proposition}
\label{cor:neutral_trap}[Popularity trap for neutrals] 
In a high-polarization event ($G_0:=G_{\min}$), fix any (even numbers of) $0<G_0<\frac{n}{3}$, $G_-=G_+=\frac{n-G_0}{2}$, and parameter ${\omega^a}\in(0,\infty)$. Then, for $\omega^p> \frac{2\omega^a G_0}{n-3G_0}$; $D^\ast>0$ holds and the interval $(D_0^{\text{high}}, D^\ast)$ is non-empty. For any misalignment cost $D>0$ satisfying $  D  \in ( D_0^{\text{high}}, D^\ast )$, the PE is the unique equilibrium and all neutral agents are strictly worse off compared to the authentic-expression benchmark.
\end{proposition}

The ``popularity trap'' creates a disconnect between individual strategic incentives and utility. In this intermediate parameter region, ($\omega^p$) is high enough to induce neutral agents to misrepresent their views, yet the cost of misalignment ($D$) is sufficiently high that the resulting increased exposure to polarized content overwhelms the utility benefits from strategic posting. The trap is self-inflicted: neutral agents individually choose to increase their payoffs taking others' posting actions as given, but in doing so, they collectively dismantle the neutral content. This forces the entire group into PE where they are worse off than if they had collectively committed to authenticity, illustrating how individually optimal actions can penalize a whole group.

Next, we focus on the most severe inefficiency: the existence of a region where PE strictly reduces the utility of \textit{all} agents relative to authenticity.

\begin{proposition}
\label{prop:everyone_worse_off}
In a high-polarization event ($G_0:=G_{min}$), fix any (even numbers of) $0<G_0<\frac{n}{3}$, $G_-=G_+=\frac{n-G_0}{2}$, and parameter $\omega^a\in (0,\infty)$. Then
there exists a threshold $\underline{\omega}_p$ such that $D^*>0$ holds, and, thus, the interval $(\max\{D_0^{\text{high}}, D_\pm^{\text{high}}\}, D^*)$ is non-empty if and only if $\omega_p>\underline{\omega}_p$. Then, for $\omega_p>\underline{\omega}_p$ and for any misalignment cost $D>0$ satisfying $D \in (\max\{D_0^{\text{high}}, D_\pm^{\text{high}}\}, D^*)$, the PE is the unique equilibrium and all agents are strictly worse off compared to the authentic-expression benchmark.
\end{proposition}

The result shows that for intense topics, the popularity trap extends to a Pareto-inferior outcome. Although opinionated agents are exposed to more aligned content in their feed, this benefit is dominated by the increased exposure to the opposing opinion with a sufficiently high distance.

Lastly, we show that all agents benefit from PE even under high-polarization if the topic intensity is low. 

\begin{proposition}
\label{prop:everyone_better_off}
In a high-polarization event ($G_0:=G_{min}$), fix any (even numbers of) $0<G_0<\frac{n}{3}$, $G_-=G_+=\frac{n-G_0}{2}$, and the parameter $\omega^a\in (0,\infty)$. Then
there exists a threshold $\overline{\omega}_p$ such that 
$
D^\ast>0
\quad\text{and}\quad D^{\mathrm{high}}_0>0$ hold, and thus,
the interval
$
\bigl(0,\min\{D^{\mathrm{high}}_0,\;D^{\mathrm{high}}_{\pm}\}\bigr)
$
is nonempty if and only if $\omega_p>\overline{\omega}_p$. Then, for $\omega_p>\overline{\omega}_p$ and for any misalignment cost $D>0$ satisfying $D \in \bigl(0,\min\{D^{\mathrm{high}}_0,\;D^{\mathrm{high}}_{\pm}\}\bigr)$, the PE is the unique equilibrium, and  all agents are strictly better off relative to the
authentic-expression benchmark.
\end{proposition}

\subsection{Low-Polarization Event}

Analogous to the high-polarization case, we first rewrite the threshold $D^{**}$. In a low-polarization event ($G_0 > n/3$), opinionated agents strategically post neutral content if and only if the misalignment cost is sufficiently low:
\[
D < D^{**} := \frac{\omega^p(3G_0-n) - \omega^a(n-G_0)}{n-G_0}.
\]
When $D < D^{**}$, PE is the unique outcome, otherwise authentic posting by all agents is the unique equilibrium.

In the PE, neutral agents are \textit{unambiguously better off} compared to the authentic benchmark for any $D > 0$. This occurs because neutral agents continue to post authentically, but now benefit from a platform where all opinionated agents strategically post the neutral view. This eliminates exposure to misaligned content and maximizes the volume of aligned posts. Consequently, the welfare analysis in this regime reduces to two possible outcomes: either a Pareto improvement (everyone is better off) or a scenario where opinionated agents are worse off while neutrals benefit.

To distinguish these cases, we define the opinionated benefit threshold $D^{\mathrm{low}}_{\pm}$. This marks the maximum misalignment cost under which opinionated agents remain better off in the PE. If $D < D^{\mathrm{low}}_{\pm}$, the popularity gains and the reduction in opposing views outweigh the cost of suppressing their true opinion. If $D > D^{\mathrm{low}}_{\pm}$, the utility losses dominate.

Proposition \ref{prop:lowpol-D-regions} fully characterizes these utility regions.

\begin{proposition}
\label{prop:lowpol-D-regions}
In a low-polarization event ($G_0:=G_{max}$), fix any (even numbers of) $G_0\in(\frac{n}{3},n)$ and
$G_- = G_+ = \frac{n-G_0}{2}$, and the parameter $\omega^a\in(0,\infty)$. $D^{\text{low}}_{\pm}< \,D^{**}(G_0)$ always holds. Then, there exists  $(\omega^p)^{**}$ and $\tilde{\omega}^p$ 
thresholds such that:

\begin{enumerate}
\item[(i)] 
$D^{**}(G_0)>0$ and
$D^{\mathrm{low}}_{\pm}>0$ hold if and only if $\omega^p>\max\{\tilde{\omega}^p,(\omega^p)^{**}\}$. 
Then, for $\omega^p>\max\{\tilde{\omega}^p,(\omega^p)^{**}\}$ and for any $D>0$ satisfying
$
0 < D < \min\{D^{\mathrm{low}}_{\pm},\,D^{**}(G_0)\}=D^{\mathrm{low}}_{\pm},
$
the PE is the unique equilibrium and all agents are strictly better off
relative to the authentic benchmark.

\item[(ii)] 
$D^{**}(G_0)>0$ holds if and only if $\omega^p>(\omega^p)^{**}$.
Then, for $\omega^p>(\omega^p)^{**}$ and for any $D>0$ satisfying
$
D^{\mathrm{low}}_{\pm} < D < D^{**}(G_0),
$
 the PE is the unique equilibrium  in which all neutral agents are strictly better off and all opinionated agents are strictly worse off compared to the authentic benchmark. 

\end{enumerate}
\end{proposition}

Proposition \ref{prop:lowpol-D-regions} establishes a fundamental asymmetry between polarization regimes: strategic incentives in low-polarization events induce \emph{convergence} toward the center rather than divergence. This structure precludes the ``everyone is worse off" region observed in the high-polarization case. Specifically, neutral agents are unambiguously better off in any popularity-driven equilibrium. By contrast, opinionated agents face a \emph{popularity trap} for sufficiently large $D$.

\subsection{Aggregate Utilities}

\begin{figure}[h!]
    \centering    \includegraphics[width=1.0\textwidth]{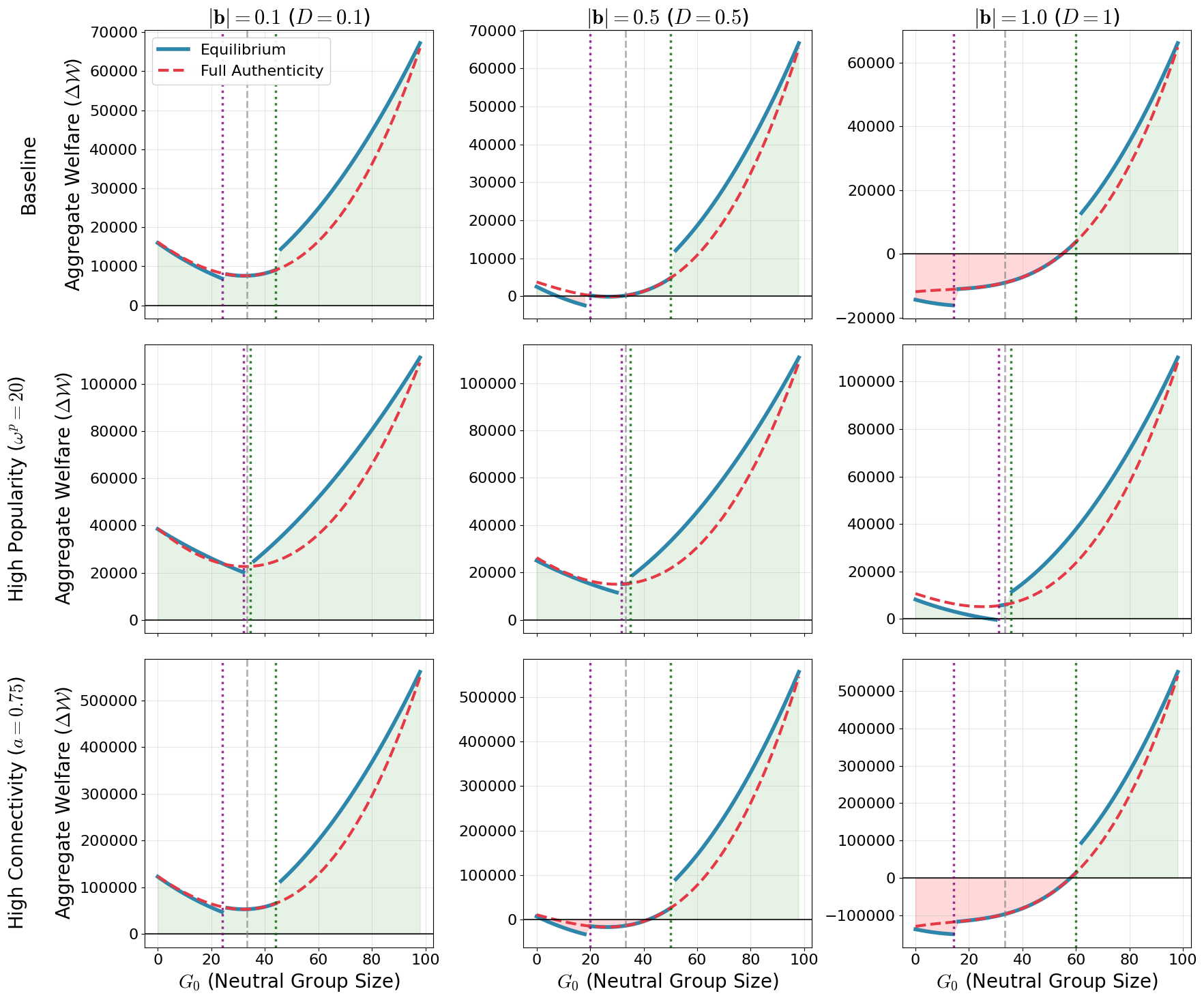}
    \caption{\footnotesize\textbf{Welfare (Aggregate Utilities) under Equilibrium and Authenticity.} 
    The figure plots aggregate welfare gains from social media interaction at equilibrium ($\Delta \mathcal{W} = \sum \Delta U_i$) relative to autarky  and authenticity benchmark outcomes. The blue solid line represents equilibrium welfare, while the red dashed line represents welfare under authentic expression.
    \textit{Notes:} 
    Red shaded regions ($\Delta \mathcal{W} < 0$) indicate parameter spaces where social media activity generates aggregate welfare losses, while green shaded regions ($\Delta \mathcal{W} > 0$) indicate aggregate welfare gains.
    Parameters: $n=100$; $a=0.25$  in Row 1 and 3, $a=0.75$ in Row 3; $\omega^p=2$ in Row 1 and 3, $\omega^p=20$ in Row 2; $\omega^a=1$; $\omega^d=1$.}
    \label{fig:welfare_level}
\end{figure}

Next, we analyze welfare, defined as the aggregate utility of all agents ($\sum_{i \in \mathcal{N}} U_i$). Proposition \ref{prop:aggregate-welfare} provides a formal result and Figure \ref{fig:welfare_level} illustrates the welfare comparison across three scenarios: autarky, authenticity, and the PE.

\begin{proposition}
\label{prop:aggregate-welfare}

\begin{itemize}

\item[(i)] In a high-polarization event ($0<G_0 < n/3$), let $\mathcal{W}^{\mathrm{auth}}$ and $\mathcal{W}^{\mathrm{eq}}$ denote the aggregate welfare under the authentic expression benchmark and under equilibrium, respectively. Then, there exists thresholds $\left(\frac{\omega^p}{\omega^a}\right)^*$, $\left(\frac{\omega^p}{\omega^a}\right)^\prime$, $D^\ast$, and $D^\prime$ such that:

\begin{itemize}
\item[(i.a)] for $\frac{\omega^p}{\omega^a}>\max\left\{\left(\frac{\omega^p}{\omega^a}\right)^*, \left(\frac{\omega^p}{\omega^a}\right)^{\prime}\right\}$; $D^*>0$ and $D' < D^\ast$, and thus, the region $(D', D^\ast)$ is non-empty. Then, for $\frac{\omega^p}{\omega^a}>\max\left\{\left(\frac{\omega^p}{\omega^a}\right)^*, \left(\frac{\omega^p}{\omega^a}\right)^{\prime}\right\}$ and for $D>0$ satisfying $D \in (D^\prime, D^\ast)$, the unique equilibrium is the PE and welfare is lower than the authenticity benchmark welfare. 

\item[(i.b)] for $\left(\frac{\omega^p}{\omega^a}\right)^*<\frac{\omega^p}{\omega^a}<\left(\frac{\omega^p}{\omega^a}\right)^{\prime}$; $D^*>0$ but $D^\prime > D^\ast$. Therefore, whenever the unique equilibrium is the PE (i.e., for $0<D<D^*$), the equilibrium is welfare-improving compared to the authentic benchmark; otherwise for $D \geq D^*$, the equilibrium is the authentic expression by all agents. 

%\item[(i.c)] for $\left(\frac{\omega^p}{\omega^a}\right)^{\prime}<\frac{\omega^p}{\omega^a}<\left(\frac{\omega^p}{\omega^a}\right)^*$; $D^*<0$ holds, and, thus, the equilibrium is the authentic expression by all agents.
\end{itemize}

\item[(ii)] In a low-polarization event ($n/3<G_0<n$), suppose that $G_0$ is the strict majority ($G_0 > n/2$). Then, the exact same results in part (i) hold under the thresholds ($\left(\frac{\omega^p}{\omega^a}\right)^{**}$, $\left(\frac{\omega^p}{\omega^a}\right)^{\prime\prime}$, $D^{\ast\ast}$) replacing the thresholds ($\left(\frac{\omega^p}{\omega^a}\right)^*$, $\left(\frac{\omega^p}{\omega^a}\right)^{\prime}$, $D^\ast$) in part (i).

\end{itemize}

\end{proposition}

The proposition highlights that whether the social media activities on the given topic helps or hurts society depends on the balance between two forces: how much users crave popularity ($\omega^p/\omega^a$) and how large is the misalignment costs.

When the topic is intermediately intense, the platform creates a ``trap'', where users chase likes rather than post their true views, but the aggregate welfare is lower compared to the authentic benchmark due to higher misaligned exposure for some agents. Conversely, when the topic intensity is low or high, this trap disappears: welfare is under the popularity-driven environment (low $D$) or agents are not involved in popularity-driven actions at all(high $D$). This implies that the PE generates welfare gains in ``light topics'' for which misalignment costs are low and the popularity-driven behavior creates positive externalities.

Figure ~\ref{fig:welfare_level} plots aggregate welfare gains ($\Delta \mathcal{W}$) under equilibrium and the authenticity benchmark relative to autarky. For high-intensity topics (Right Panels, $|b|=1.0$), we observe a prolonged net welfare loss region unless $\omega^p$ is large (Row 2). Conversely, for low-intensity topics (Left Panels, $|b|=0.1$), the platform generates substantial welfare gains (green region), and particularly in the low-polarization regime (right side of panels). This stark contrast confirms that social media is neither inherently beneficial nor harmful; rather, its welfare impact is state-dependent. 

Besides, Figure ~\ref{fig:welfare_level} shows that the network density has distinct implications on welfare, even though it has no effect on posting decisions in our model. Comparing the values at the y-axes of Rows 1 and 3, with parameters $a=0.25$ and $a=0.75$ all else is fixed, higher density magnifies welfare gains when social media interaction is welfare-improving, and amplifies welfare losses when it is welfare-reducing.

\section{An Algorithmic Tool: Preference-Based Exposure}
\label{sec:pvm}

In the final part of the paper, we study the role of a widely used platform algorithm through which users are selectively shown content matching their preferences. We refer to it as \emph{Post–Viewer Match (PVM)} algorithm.

Before analyzing the platform's active role, consider a benchmark case where the underlying social network is fully homophilic, such that every agent $i$'s follower set $\mathcal{A}_i$ consists exclusively of individuals sharing the same authentic opinion with agent $i$. In this setting, agents cannot expand their audience by adopting a different stance, as their potential viewers are exogenously restricted to like-minded people. Consequently, popularity-driven strategic incentives vanish, equilibrium posting remains authentic, and welfare outcomes correspond to the authentic benchmark derived earlier. However, modern platforms frequently distribute content beyond direct follower networks, algorithmically determining exposure based on user interests. We therefore introduce a platform-driven exposure rule that imposes homophily by matching the specific content of a post to the authentic preferences of potential viewers, which allows us to expand the friendship-based homophily case to an algorithmic homophily.

\paragraph{Post--viewer--match algorithm (PVM).}

For the post created by any agent $i \in \mathcal{N}$, the platform selects a visibility set $\mathcal{V}_i \subseteq \mathcal{N}$, that is the set of users to whom $i$'s post is displayed in their feeds, which is not necessarily based on the follower set $\mathcal{A}_i$, but can be solely decided by the platform algorithm. The number of users in this set is $V_i = |\mathcal{V}_i| \leq V^*_i$, where $V^*_i\leq n$ is a capacity level for visibility of agent $i$'s post. A reasoning for imposing a capacity is that users spend limited time on social media, and since our model abstracts from the determinants of time spent, we capture limited user attention simply through an exogenous visibility cap parameter.

Next, we revisit the timing of the model. Posts are created at time $t=0$, and at this step, agents are assumed to know the platform’s algorithmic rule\footnote{While the assumption 
that agents know the algorithm may appear strong, it is consistent with the mechanics of \textit{like-based 
suggestions} widely used by social media platforms (\cite{bakshy2015exposure}).  
 This feedback---where likes reveal users’ preferences and the platform adjusts visibility in response---illustrates how platforms gradually learn user types. On platforms that use \textit{like-based} suggestions, it is therefore reasonable for agents to expect the algorithm to behave like a post-viewer-match rule, showing posts primarily to users whose opinions align with the content.}. After posts are created, the platform chooses the set $\mathcal{V}_i$ for each $i \in \mathcal{N}$ at time $t=1$, before making these posts visible on social media at time $t=2$. The information structure and the timing of posting actions and reactions to posts are same as in Section \ref{model}. 

\paragraph{Platform's maximization at time $t=1$.} Given post $c_i$, the platform chooses the set $\mathcal{V}_i$ for each $i$ that maximizes the number of likes subject to the visibility constraint $V_i \leq V^\ast_i$ (referring to engagement maximization by platform):
\[
\max_{\mathcal{V}_i\subseteq \mathcal{N}}\; \sum_{j\in \mathcal{V}_i} r_{ji}
\quad\text{s.t.}\quad V_i \leq V^\ast_i.
\]

The analysis below compares two distinct algorithms, called as the \textit{representative algorithm}---that captures the representative setup used so far in the article---and \textit{post-viewer match algorithm}. 

\medskip

\textit{Representative algorithm (RA)}. This algorithm exactly matches the representative exposure, but under a visibility cap. Under this algorithm, each agent $i$'s post is shown to $V_i \cdot \frac{G_{b^\ast}}{n}$ number of agents from each opinion group $b^\ast \in \mathcal{O}$.\footnote{We assume that parameters in $V_i \cdot \frac{G_{b^\ast}}{n}$ satisfy conditions such that there is no divisibility issue.} Consequently, the total likes of agent $i$ is $R_i = V_i \cdot \frac{G_{c_i}}{n}$, where $G_{c_i}$ is the group size of agents in society holding the view that is same as $c_i$. This implies the platform maximizes the number of likes by setting $V_i=V^\ast_i$, and hence, total likes for each agent $i$ is: 

\[
R_i = V^\ast_i \cdot \frac{G_{c_i}}{n}.
\]

\medskip 

\textit{Post-viewer-match algorithm (PVM)}. Under this algorithm, the visibility of each post is determined by the exact opinion matching between the post and viewers. Therefore, an optimal $\mathcal{V}_i$ fills the exposure set with users whose authentic opinions match post $c_i$, up to the availability of such users in $\mathcal{N}$. Beyond this level, since there would be no additional likes from other types, we simply assume that posts are not shown to any other agents holding different opinions. Formally, agent $i$’s content is displayed to an agent $j$ only if $c_i = b_j$. Consequently, under post-viewer-match algorithm, $R_i = \min\{V_i,  G_{c_i}\}$, implying that: 

\[
R_i \;=\; \min\{V^\ast_i,\; G_{c_i}\}.
\]

We now study the implications of these two algorithms on polarization under homogeneous visibility caps
and utility parameters across agents.

\begin{proposition}
\label{proposition:algorithm}
\begin{itemize}
Suppose the baseline utility, utility parameters, and visibility caps are homogeneous: $
H_i=\overline{H},\quad
\omega_i^p=\omega^p,\quad \omega_i^a=\omega^a,\quad \omega_i^d=\omega^d,\quad
a_i=a,\quad
 V_i^*=k\quad \forall i\in\mathcal{N}.$ 

\item[(i)] In a high-polarization event $(0<G_0 < n/3)$, the social media under RA is weakly more polarized than social media under PVM ($C_0^{RA} \leq C_0^{PVM}$), and strictly more polarized ($C_0^{RA} < C_0^{PVM}$) if and only if:
    \begin{equation*} 
    G_0 < G^\ast = \frac{(\omega^p-\omega^d |b|) n}{3 \omega^p + 2 \omega^a - \omega^d |b|} 
    \quad \text{and} \quad 
    k \leq G_0 \frac{\omega^p+\omega^a}{\omega^p-\omega^d |b|}.
    \end{equation*}

    \item[(ii)] In a low-polarization event $(n/3<G_0 < n)$,
    the social media under RA is weakly less polarized than social media under PVM ($C_0^{RA} \leq C_0^{PVM}$), and strictly less polarized ($C_0^{RA} < C_0^{PVM}$) if and only if:
    \begin{equation*}   
    G_0 > G^{\ast\ast} = \frac{(\omega^p+\omega^a)n}{3\omega^p -2 \omega^d |b| +\omega^a} 
    \quad \text{and} \quad  
    k \leq \frac{n-G_0}{2}  \frac{\omega^p+\omega^a}{\omega^p-\omega^d|b|}.
    \end{equation*}
\end{itemize}
\end{proposition}

\begin{figure}[h!]
    \centering
    \includegraphics[width=0.6\textwidth]{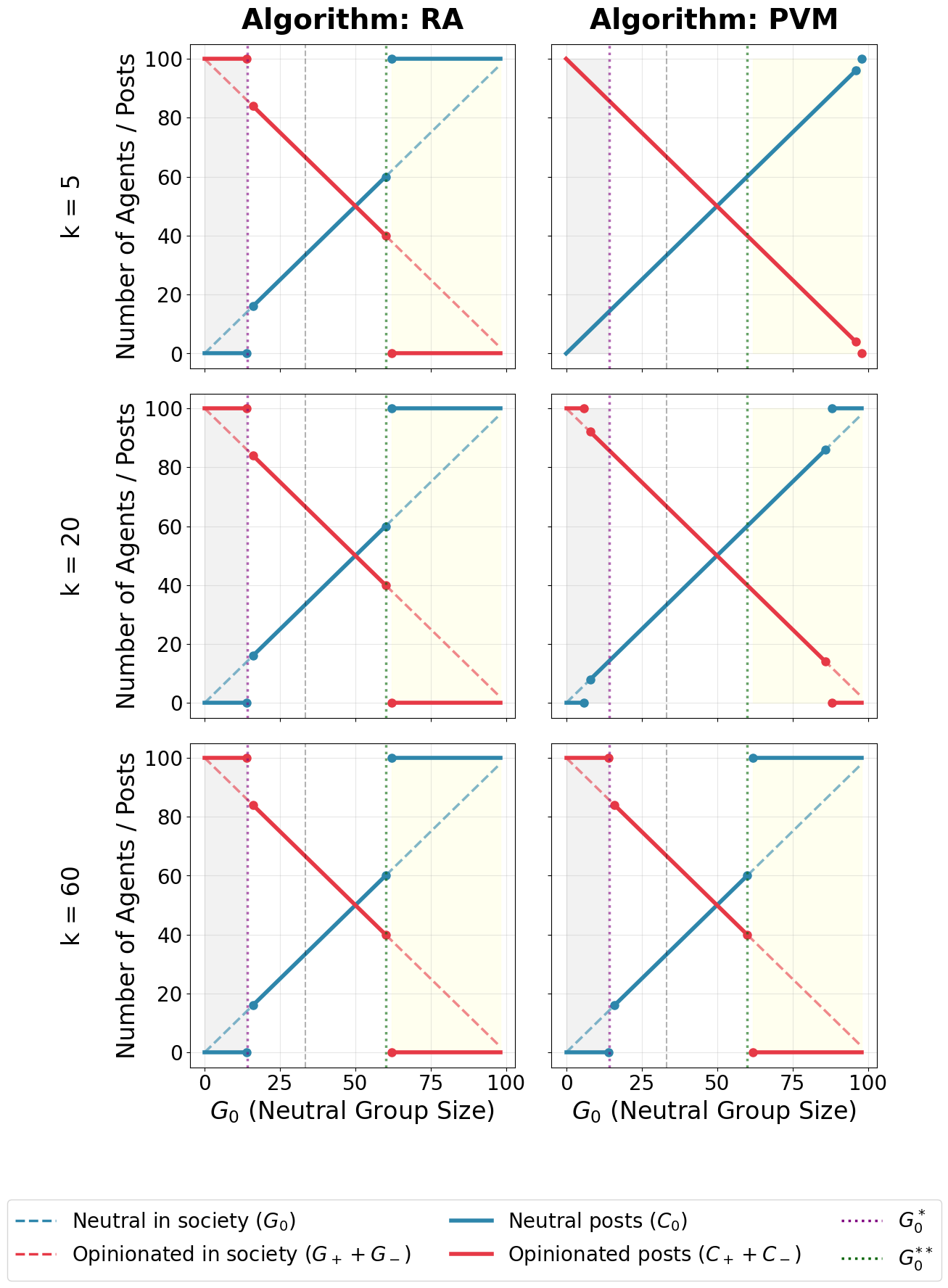}
    \caption{Polarization Under RA versus PVM}
    \label{fig:PVM_vs_RA}
    \vspace{0.3em}
    \begin{minipage}{0.98\textwidth}
    \footnotesize
    \textbf{Notes:} \footnotesize The figure illustrates equilibrium posting behavior (solid lines) compared to the authentic opinion distribution (dashed lines) under two algorithmic regimes: the Representative Algorithm (RA, left column) and the Post-Viewer-Match algorithm (PVM, right column). 
    The rows correspond to three distinct visibility caps: very low ($k=5$, top), low ($k=20$, middle), and high ($k=60$, bottom).
    Parameters are set to $n=100$, $\omega^p=2$, $\omega^a=1$, and $\omega^d|b|=1$.
    Under RA, posting incentives are independent of $k$. Under PVM, tight visibility constraints ($k=5, 20$) narrow the region of strategic deviation in high-polarization events, while relaxed constraints ($k=60$) bring back the popularity-driven posting at equilibrium.
    \end{minipage}
\end{figure}

Proposition~\ref{proposition:algorithm} highlights that algorithmic targeting can discipline popularity-driven behavior, but crucially depending on the level of visibility. The key mechanism is that post--viewer matching restricts exposure to like-minded audiences, thereby weakening incentives to strategically distort content toward more popular opinions if visibility is sufficiently limited.

When society is initially polarized and the neutral group is small, under post--viewer matching, distortions are eliminated when visibility is sufficiently small, because the cap limits the potential additional likes from adopting a more popular stance. However, representative algorithm still incentivizes neutrals to adopt opinionated stances.

\noindent Figure \ref{fig:PVM_vs_RA} shows the differences in polarization levels under RA and PVM, clarifying that PVM narrows the strategic posting region. Consequently, PVM restores authentic expression in certain cases.

Contrary to the common concern that homophilic exposure increases polarization, preference-based matching---an algorithmic homophily---can reduce polarization by limiting the space for strategic expression when visibility of posts are sufficiently small.

From a policy perspective, the analysis reveals an important tradeoff. Platform design choices that increase visibility—such as extended feeds or short-form videos—can alleviate the polarization-amplifier role of popularity incentives, whereas limiting visibility under the PVM algorithm can overturn these effects. Conversely, binding caps reduce the incentive for moderation during unified events, and thus, reduce the welfare gains from it.

\section{Conclusion}
\label{sec:conclusion}

This paper develops a utilitarian framework to analyze strategic expression on social media. By specifying utility over posting, exposure, and engagement, the framework fills a critical gap in the literature, providing a microfoundation for individual utility and aggregate welfare that maps to directly observable platform metrics. 

In the model, agents derive utility from popularity of their posts and exposure to aligned/ misaligned content. This approach introduces a novel channel of distortion of utilities through strategic opinion expression—separating it from canonical models of belief updating, persuasion, learning, and transmission of (mis)information.

Our findings show that social media activities affect individual utilities in a heterogeneous way, and popularity-driven behavior distorts expressed opinions online. A central implication is that these activities as well strategic distortions either redistributes or magnifies the welfare (aggregate utilities) in a state-dependent way. 

Moreover, we identify a phenomenon called ``popularity trap"--- a source of coordination failure: although posting a more popular opinion is individually optimal, collective strategic posting by similar agents eliminates those agents' authentic viewpoint from the platform, increasing their exposure to misaligned content and eventually reducing their individual utilities that could be prevented under coordinated authentic expression.

Lastly, our findings show that homophilic exposure or algorithmic design by platforms discipline popularity-driven behavior, and hence, narrow the region for popularity traps and limiting its welfare effects. This means contrary to the widely accepted view that homophily increases polarization, our findings present a channel in which homophily overturns the impact of strategic distortion on polarization.

These findings have direct implications for platform design and policy. Interventions that uniformly promote engagement or popularity may be beneficial in low-stakes environments, yet harmful in polarized and high-intensity contexts. Optimal policy is therefore inherently state-dependent. Mechanisms that dampen popularity incentives during intense or divisive events may mitigate popularity traps and welfare losses, while also eliminating the benefits of social media interaction in some other context.

\begin{singlespace}
\printbibliography    
\end{singlespace}

\newpage

\section*{Appendix}
\label{sec:appendix}

\subsubsection*{Proof of Lemma \ref{prop:SPNE}}

\begin{proof}
Fix an equilibrium $(\mathbf c,\mathbf R)$ and an arbitrary agent $i \in \mathcal N$, and solve for the SPNE using backward induction.

\paragraph{Reaction stage $t=1$.}
Fix $\mathbf c$ and consider any follower $j\in\mathcal A_i$ who observes $c_i$ and chooses $r_{ji}\in\{0,1\}$. Since $r_{ji}$ enters $U_j$ only through $R_i$, setting $r_{ji}=1$ increases $R_i$ by one unit. From \eqref{utilityf}, the induced change in $U_j$ is
\[
U_j(r_{ji}=1\mid \mathbf c,\mathbf R_{-ji})-U_j(r_{ji}=0\mid \mathbf c,\mathbf R_{-ji})
=
\omega_j^a \mathbf 1\{c_i=b_j\}-\omega_j^d \mathbf 1\{c_i\neq b_j\}\,|b_j-c_i|.
\]
where $\omega_j^a>0$ and $\omega_j^d>0$. Therefore, for $\omega_j^a>0$ and $\omega_j^d>0$, the unique best response is
\[
r_{ji}^*=1_{\{c_i=b_j\}} \qquad \text{and} \qquad r_{ji}^*=0_{\{c_i \neq b_j\}}.
\]
Therefore, given $c_i$, agent $i$'s equilibrium level of popularity is
\[
R_i=\sum_{j\in\mathcal A_i} r_{ji}^*
=\sum_{j\in\mathcal A_i}1_{\{b_j=c_i\}}.
\]
In particular, if $c_i\notin\mathcal O$, then $r_{ji}^*=0_{\{c_i \neq b_j\}}$ for all $j$, hence $R_i=0$.

\paragraph{Posting stage $t=0$.}
Fix the strategy profile of all agents other than $i$, and write agent $i$'s utility as
\[
U_i
=
H_i
+\omega_i^p R_i
+\omega_i^a \sum_{j\in \mathcal{N}_i}\mathbf 1\{c_j=b_i\}R_j
-\omega_i^d \sum_{j\in \mathcal{N}_i}\mathbf 1\{c_j\neq b_i\}R_j|b_i-c_j|.
\]
Let $F_i$ collect all terms in $U_i$ that excludes all utility gains and losses from self-posting (i.e., terms involving only $\mathbf c_{-i}$ and $\mathbf R_{-i}$). Since $i\in \mathcal{N}_i$ (self-exposure), we can decompose the two exposure sums into the $j=i$ term and the $j\neq i$ terms, yielding
\[
U_i=H_i + F_i
+\omega_i^p R_i
+\omega_i^a \mathbf 1\{c_i=b_i\}R_i
-\omega_i^d \mathbf 1\{c_i\neq b_i\}R_i\,|b_i-c_i|.
\]
Equivalently,
\[
U_i=H_i + F_i + R_i\Big(\omega_i^p+\omega_i^a \mathbf 1\{c_i=b_i\}-\omega_i^d \mathbf 1\{c_i\neq b_i\}|b_i-c_i|\Big).
\]

Now take any $c_i\notin\mathcal O$. By Step 1, $R_i=0$, so $U_i=F_i$.
Next consider $c_i=b_i\in\mathcal O$. Then,
\[
R_i=\sum_{j\in\mathcal A_i}\mathbf 1\{b_j=b_i\}=A_{i,b_i} > 0,
\]
and, hence,
\[
U_{i,\{c_i =b_i\}}=H_i + F_i + A_{i,b_i}(\omega_i^p+\omega_i^a) > H_i + F_i=U_{i, \{c_i\notin\mathcal O\}}.
\]
Thus, posting $c_i=b_i$ dominates posting any $c_i\notin\mathcal O$. Consequently, any maximizer of $U_i(\cdot)$ is selected from $\mathcal O$.
Therefore, in equilibrium $c_i\in\mathcal O$ for all $i\in\mathcal N$.

\end{proof}

\subsubsection*{Proof of Proposition \ref{prop1}}
\begin{proof}
We characterize equilibrium posting behavior in the three-opinion environment
$\mathcal{O}=\{-b,0,+b\}$ under the symmetric configuration
$G_- = G_+ = (n-G_0)/2$ and representative social media, where
$A_{i,b^\ast}=a_i G_{b^\ast}$ for all $i\in\mathcal N$ and $b^\ast\in\mathcal O$.

\paragraph{(i) High-polarization event: $G_0<n/3$.}
In this case $G_0<G_-=G_+=(n-G_0)/2$.

\smallskip
\noindent\emph{Neutral agents.}
Fix $i\in\mathcal G_0$. Deviations to $+b$ and $-b$ yield identical payoffs, so $i$
posts $0$ if and only if
\[
(\omega_i^p+\omega_i^a)a_iG_0
\;\ge\;
a_i\frac{n-G_0}{2}(\omega_i^p-\omega_i^d|b|).
\]
Rearranging gives
\[
a_iG_0\big(3\omega_i^p+2\omega_i^a-\omega_i^d|b|\big)
\;\ge\;
a_in(\omega_i^p-\omega_i^d|b|),
\]
which is equivalent to $G_0\ge G_0^\ast$ ($G_0^\ast$ is an individual-specific threshold), where
\[
G_0^\ast
=
\frac{(\omega_i^p-\omega_i^d|b|)\,n}{3\omega_i^p-\omega_i^d|b|+2\omega_i^a}.
\]
Thus, if $\omega_i^p-\omega_i^d|b|>0$ and $G_0<G_0^\ast$, the neutral agent $i$ strictly prefer
to post an opinionated content. Under the even-split convention, $C_+=C_->G_+=G_-$ and
$C_0<G_0$. Hence, social media is strictly more polarized than society. If
$G_0\ge G_0^\ast$ holds for all netural agents, then all neutral agents post authentically and polarization is equal in these two cases. This threshold can be rewritten by using the individual popularity parameter $\omega^p_i$ as follows:

$$\omega^p_i > \frac{2 \omega^a_i G_0 + \omega^d_i |b| (n - G_0)}{n - 3 G_0}.$$

\smallskip
\noindent\emph{Opinionated agents.}
Fix $i$ with $b_i=+b$ (the case $b_i=-b$ is symmetric). For a $+b$ type agent, posting a neutral content yields
$a_iG_0(\omega_i^p-\omega_i^d|b|)$, while posting $-b$ yields
$a_iG_-(\omega_i^p-\omega_i^d|2b|)$. Since $G_->G_0$ and $|2b|>|b|$, both deviations yield
a strictly lower payoff than posting $+b$ ($a_iG_+(\omega^p_i +\omega^a_i)$). Hence, opinionated agents post authentically at equilibrium.

\paragraph{(ii) Low-polarization event: $G_0>n/3$.}
In this case $G_0>G_-=G_+=(n-G_0)/2$. By symmetry, it suffices to consider agents with
$b_i=+b$.

\smallskip
\noindent\emph{Opinionated agents.}
Posting $+b$ is optimal if and only if
\[
(\omega_i^p+\omega_i^a)a_iG_+
\;\ge\;\max\{
a_iG_0(\omega_i^p-\omega_i^d|b|), a_i G_-(\omega^p_i -2\omega_i^d|b|)\}  = a_iG_0(\omega_i^p-\omega_i^d|b|).
\]
Substituting $G_+=(n-G_0)/2$ and rearranging yields $G_0\le G_0^{\ast\ast}$, where
\[
G_0^{\ast\ast}
=
\frac{(\omega_i^p+\omega_i^a)\,n}{3\omega_i^p-2\omega_i^d|b|+\omega_i^a}.
\]

The threshold can be rewritten in individual popularity parameter $\omega^p_i$ as follows:

   $$\omega^p_i > \frac{\omega^a_i (n - G_0) + 2 \omega^d_i |b| G_0}{3 G_0 - n}.$$

If $G_0>G_0^{\ast\ast}$ and $\omega_i^p-\omega_i^d|b|>0$, opinionated agents $i$ strictly
prefer to deviate to $0$, implying $C_0>G_0$ and $C_+=C_-<G_+=G_-$. Social media is then
strictly less polarized. If $G_0\le G_0^{\ast\ast}$, opinionated agents post
authentically and polarization is equal in both cases.

\smallskip
\noindent\emph{Neutral agents.}
For $i\in\mathcal G_0$, posting $0$ yields $(\omega_i^p+\omega_i^a)a_i G_0$, while deviating
to $\pm b$ yields $a_i (n-G_0)/2(\omega_i^p-\omega_i^d|b|)$. Since $G_0>(n-G_0)/2$ and
$\omega_i^a>0$, posting $0$ strictly dominates any other posting decision. Neutral agents post authentically at equilibrium.
\end{proof}

\subsubsection*{Proof of Corollary ~\ref{unification}}
\begin{proof}
Consider the three-opinion environment of Proposition~\ref{prop1} with homogeneous
parameters $(\omega^p,\omega^a,\omega^d)$ and assume $\omega^p>\omega^d|b|$.

By Proposition~\ref{prop1} (i) in a high-polarization event
($G_0=G_{\min}$), all neutral agents post opinionated content if and only if
$
G_0 < G^{\mathrm{neutral}} = G^\ast 
=
\frac{(\omega^p-\omega^d|b|)n}{3\omega^p-\omega^d|b|+2\omega^a}.
$
By Proposition~\ref{prop1} (ii) in a low-polarization event
($G_0=G_{\max}$), all opinionated agents post neutral content if and only if
$
G_0 > G^{\ast\ast}
=
\frac{(\omega^p+\omega^a)n}{3\omega^p-2\omega^d|b|+\omega^a}.
$
Since $G_-+G_+=n-G_0$, this condition is equivalent to
$G_-+G_+<n-G^{\ast\ast}$. Define
$
G^{\mathrm{opin}}:=n-G^{\ast\ast}.
$

We now compare the two thresholds $G^{\mathrm{neutral}}$ and
$G^{\mathrm{opin}}$. Dividing by $n>0$, it suffices to show that
\[
\frac{\omega^p-\omega^d|b|}{3\omega^p-\omega^d|b|+2\omega^a}
<
\frac{2(\omega^p-\omega^d|b|)}{3\omega^p-2\omega^d|b|+\omega^a}.
\]
In the parameter region where deviations occur,
$\omega^p-\omega^d|b|>0$, so the common factor cancels. Both denominators are
positive, and cross-multiplication yields
\[
3\omega^p-2\omega^d|b|+\omega^a
<
2(3\omega^p-\omega^d|b|+2\omega^a)
=
6\omega^p-2\omega^d|b|+4\omega^a,
\]
which simplifies to $0<3\omega^p+3\omega^a$. This inequality holds since
$\omega^p,\omega^a>0$.

Therefore,
\[
G^{\mathrm{neutral}}<G^{\mathrm{opin}}.
\]
\end{proof}

\subsubsection*{Proof of Proposition \ref{cor:neutral_trap}}

Fix a high-polarization event with $G_0<\frac{n}{3}$ and 
$G_- = G_+ = \frac{n-G_0}{2}$. Impose homogeneity and define 
$D := \omega^d |b| \in (0,\infty)$. 

\paragraph{Neutral agents behavior}
Fix a neutral agent $i \in \mathcal G_0$ with $b_i = 0$.
We compare her utility from posting $c_i=0$ with that from posting 
$c_i \in \{-b,+b\}$.

\smallskip
The individual posting decision is obtained by comparing the payoff from authentic
posting with that from an opinionated post as shown in Proposition \ref{prop1}:
\[
(\omega^p+\omega^a) a G_0 
< 
(\omega^p-D)a \frac{n-G_0}{2}.
\]
This inequality is equivalent to
\[
D < D^\ast 
:= 
\frac{\omega^p (n-3G_0) - 2\omega^a G_0}{n-G_0}.
\]

For $D^\ast>0$ to hold, we need
$\omega^p>\frac{2\omega^a G_0}{\,n-3G_0\,}.$ 

Suppose $0<D < D^\ast$ holds. 
 
\paragraph{Utilities.}
We now compute the utilities of a neutral agent under the two posting choices.

\smallskip
\noindent\emph{(a) If $i$ posts $c_i=\pm b$.}
Then $R_i = aG_{\pm} = a\frac{n-G_0}{2}$. In this case, all posts observed by $i$
are opinionated and lie at distance $|b|$ from $0$. Each such post has popularity
$a\frac{n-G_0}{2}$, so total misaligned exposure equals
$(an+1)\cdot a\frac{n-G_0}{2}$. Hence
\begin{equation}
\label{eq:neutral-dev-utility}
U_0^{eq}=\overline{H}
+
\omega^p a\frac{n-G_0}{2}
-
D (an+1)a\frac{n-G_0}{2}.
\end{equation}

\smallskip
\noindent\emph{(b) If $i$ posts $c_i=0$.}
Then $R_i = aG_0$. Among the $an$ posts from others in $\mathcal{N}_i$,
a fraction $G_0/n$ are neutral and a fraction $(n-G_0)/n$ are opinionated. Agent $i$
therefore observes $aG_0$ neutral posts from others plus her own post, yielding
$(aG_0+1)$ aligned posts, each with popularity $aG_0$. Aligned exposure is thus
\[
S_i^{0} = (aG_0+1)aG_0.
\]
In addition, $i$ observes $aG_+=a\frac{n-G_0}{2}$ posts of each opinionated type,
each with popularity $a\frac{n-G_0}{2}$ and distance $|b|$. Total misaligned exposure is
\[
S_i^{+}+S_i^{-}
=
a^2(G_+^2+G_-^2)
=
a^2\frac{(n-G_0)^2}{2}.
\]
Therefore,
\begin{equation}
\label{eq:neutral-auth-utility}
U_i^{\text{auth}} =\overline{H}
+
\omega^p aG_0
+
\omega^a (aG_0+1)aG_0
-
D a^2\frac{(n-G_0)^2}{2}.
\end{equation}

We define the neutral benefit threshold $D_0^{\text{high}}$ as the critical misalignment cost at which a neutral agent is indifferent between the popularity-driven equilibrium and the authentic benchmark (i.e., satisfying $U_0^{eq} = U_i^{\text{auth}}$).

Equating the utility expressions derived in \eqref{eq:neutral-dev-utility} and \eqref{eq:neutral-auth-utility}:
\[
\omega^p a\frac{n-G_0}{2} - D (an+1)a\frac{n-G_0}{2}
=
\omega^p aG_0 + \omega^a (aG_0+1)aG_0 - D a^2\frac{(n-G_0)^2}{2}.
\]
Rearranging terms to group components involving $D$ on the right-hand side and the remaining terms on the left:
\[
\omega^p a \left( \frac{n-G_0}{2} - G_0 \right) - \omega^a aG_0(aG_0+1)
=
D a \frac{n-G_0}{2} \left[ (an+1) - a(n-G_0) \right].
\]

Solving for $D$ yields the threshold definition:
\[
D_0^{\text{high}}
:=
\frac{\omega^p (n-3G_0) - 2\omega^a G_0(aG_0+1)}
{(n-G_0)(aG_0+1)}.
\]

Consequently, when the actual misalignment cost exceeds this threshold ($D > D_0^{\text{high}}$), the popularity gains are insufficient to offset the increased polarization, leaving neutral agents strictly worse off in the popularity-driven outcome compared to the authentic benchmark. 

\paragraph{Comparison.}
Subtracting \eqref{eq:neutral-auth-utility} from \eqref{eq:neutral-dev-utility} and simplifying,
neutral agents are strictly worse off under the popularity-driven equilibrium, if achieved, than under
authentic expression if and only if
\[
D_0^{\text{high}} < D < D^\ast.
\]

Lastly, given that $aG_0>0$, the strict ordering of $D_0^{\text{high}}<D^*$ always holds:

\[
D_0^{\text{high}}
=
\frac{\omega^p (n-3G_0) - 2\omega^a G_0(aG_0+1)}
{(n-G_0)(aG_0+1)}
<
D^\ast= \frac{\omega^p (n-3G_0) - 2\omega^a G_0}{n-G_0}.
\]
 
For $\omega^p>\frac{2\omega^a G_0}{\,n-3G_0\,}$, we have $D^\ast>0$. This establishes the existence of an intermediate range of misalignment costs $D_0^{\text{high}} < D < D^\ast$, and for $D>0$ at this range,
each neutral agent post opinionated yet are worse off at equilibrium compared to the authentic expression benchmark.

\subsubsection*{Proof of Proposition \ref{prop:everyone_worse_off}}

\begin{proof}
 Proposition \ref{cor:neutral_trap} shows the existence of $D$ region for neutral agents being worse off at the unique popularity-driven equilibrium compared to the authenticity benchmark. First, we show such a region exists for opinionated agents, too. These together would imply there exists a region in which all agents are worse off.  

Suppose $D<D^*$, so the unique equilibrium is popularity-driven.

\smallskip
\noindent\emph{ Opinionated agents.}
Fix $i$ with $b_i=+b$ (the case $b_i=-b$ is symmetric). Under this case all opinionated agents post their authentic view (as shown in Proposition \ref{prop1}), then, $R_i=a\frac{n-G_0}{2}$.
By even split, among the $an$ posts from others in $N_i$, half are $+b$ and half are $-b$.
Including self-exposure, $i$ observes $(\frac{an}{2}+1)$ aligned posts and $\frac{an}{2}$ misaligned posts,
each with popularity $a\frac{n-G_0}{2}$. Hence
\[
S_i^{b_i}=\Big(\frac{an}{2}+1\Big)a\frac{n-G_0}{2},
\qquad
S_i^{-b_i}=\Big(\frac{an}{2}\Big)a\frac{n-G_0}{2}.
\]
Misalignment is at distance $2|b|$, so the misalignment-exposure-based utility loss is $2D\,S_i^{-b_i}$.
Therefore
\[
 U^{eq}_\pm =\overline{H}
+
\omega^p a\frac{n-G_0}{2}
+\omega^a\Big(\frac{an}{2}+1\Big)a\frac{n-G_0}{2}
-2D\Big(\frac{an}{2}\Big)a\frac{n-G_0}{2}.
\]

The utility of an opinionated agent under the authenticity benchmark is:

\[U_{i,\pm}^{\text{auth}} =\overline{H} +\omega^p a \frac{n-G_0}{2}+ \omega^a (a\frac{n-G_0}{2} +1)a\frac{n-G_0}{2} - D\left((aG_0)^2 + 2\left(a\frac{n-G_0}{2}\right)^2\right)\]

The difference in utility for these agents between the equilibrium and the authentic benchmark is given by:
\begin{align*}
    U^{eq}_\pm - U_{i,\pm}^{\text{auth}}
    &= \left[ \omega^a\left(\frac{an}{2}+1\right)A - 2D\left(\frac{an}{2}\right)A \right]
    - \left[ \omega^a(A+1)A - D\left((aG_0)^2 +2A^2\right) \right] \\
    &= \omega^a A \left[ \left(\frac{an}{2} + 1\right) - (A + 1) \right]
    + D \left[ (aG_0)^2 + 2A^2 - 2\left(\frac{an}{2}\right)A \right]
\end{align*}
where $A = a\frac{n-G_0}{2}$. We simplify the terms associated with $\omega^a$ and $D$ separately.

For the $\omega^a$ term:
\begin{align*}
    \omega^a A \left( \frac{an}{2} - A \right)
    &= \omega^a \left( a\frac{n-G_0}{2} \right) \left( \frac{an}{2} - \frac{an - aG_0}{2} \right) \\
    &= \omega^a \left( a\frac{n-G_0}{2} \right) \left( \frac{aG_0}{2} \right) \\
    &= \frac{a^2}{4} \omega^a G_0 (n - G_0).
\end{align*}

For the $D$ term:
\begin{align*}
    D \left[ (aG_0)^2 + 2A^2 - an A \right]
    &= D \left[ a^2 G_0^2 + 2 \frac{a^2(n-G_0)^2}{4} - an \frac{a(n-G_0)}{2} \right] \\
    &= \frac{a^2 D}{4} \left[ 4G_0^2 + 2(n-G_0)^2 - 2n(n-G_0) \right] \\
    &= \frac{a^2 D}{4} \left[ 4G_0^2 + 2(n^2 - 2nG_0 + G_0^2) - (2n^2 - 2nG_0) \right] \\
    &= \frac{a^2 D}{4} \left[ 6G_0^2 - 2nG_0 \right].
\end{align*}

Combining these components yields the final expression:
\[
    U^{eq}_\pm - U_{i,\pm}^{\text{auth}} = \frac{a^2}{4} \left[ \omega^a G_0(n-G_0) + D(6G_0^2 - 2nG_0) \right].
\]

For any $G_0<n/3$, it always holds that $6G_0^2 < 2nG_0$. Then,

\[D^{\text{high}}_\pm = \frac{ \omega^a G_0(n-G_0)}{(2nG_0-6G_0^2 )}>0,\]

where $U^{eq}_\pm<U^{auth}_\pm$ at this equilibrium if and only if $D>D_\pm^{\text{high}}$ holds.

Lastly, we demonstrate that the region where opinionated agents are worse off is compatible with the existence of the popularity-driven equilibrium; that is, we show the interval $(D_\pm^{\text{high}}, D^*)$ is non-empty for a valid range of parameters. We require
\[
D_\pm^{\text{high}} < D^*=\frac{\omega^p (n-3G_0) - 2\omega^a G_0}{n-G_0}.
\]
Substituting the derived expressions:
\[
\frac{ \omega^a G_0(n-G_0)}{2G_0(n-3G_0)} < \frac{\omega^p (n-3G_0) - 2\omega^a G_0}{n-G_0}.
\]
Notice that the benefit threshold $D_\pm^{\text{high}}$ is independent of the popularity weight $\omega^p$, whereas the existence threshold $D^*$ is strictly increasing in $\omega^p$ (since $n > 3G_0$ implies $n-3G_0 > 0$). Therefore, for any fixed opinion distribution $G_0 < n/3$ and alignment weight $\omega^a$, there exists a sufficiently large popularity weight $\omega^p$ such that $D^* > D_\pm^{\text{high}}$. Specifically, rearranging the inequality, this condition holds whenever:
\[
\omega^p > 
 \frac{1}{n-3G_0} \left[ (n-G_0) D_\pm^{\text{high}} + 2\omega^a G_0 \right].
\]

From Proposition \ref{cor:neutral_trap}, $D^*$ is positive for $\omega^p>\frac{2\omega^aG_0}{n-3G_0}$.

Then, we end up with:

\[
\underline{\omega}^p
= \max\left\{\frac{1}{n-3G_0} \left[ (n-G_0) D_\pm^{\text{high}} + 2\omega^a G_0 \right], \frac{2\omega^aG_0}{n-3G_0}\right\}.
\]

For $\omega^p> \underline{\omega}^p$, the set of misalignment costs satisfying $\max\{D_0^{\text{high}}, D_\pm^{\text{high}}\} < D < D^\ast$,  is non-empty. Choosing $D>0$ within this interval ensures that the popularity-driven equilibrium is unique and yields strictly lower utility for both neutral and opinionated agents compared to the authentic benchmark.

\end{proof}

\subsubsection*{Proof of Proposition \ref{prop:everyone_better_off}}

\begin{proof}
Fix a high-polarization event with $G_0<\frac{n}{3}$ and
$G_-=G_+=\frac{n-G_0}{2}$. Let $D:=\omega^d|b|$.

 The neutral-type
utility comparison derived in Proposition~\ref{cor:neutral_trap} yields
\[
\Delta U^0 := U^{\mathrm{eq},0}-U^{\mathrm{auth},0} > 0
\quad\Longleftrightarrow\quad
D < D^{\mathrm{high}}_0 <D^\ast
\]

Proposition~\ref{prop:everyone_worse_off} shows that the condition for opinionated agents to be strictly better off: 

\[
\Delta U^\pm := U^{\mathrm{eq},\pm}-U^{\mathrm{auth},\pm} > 0
\quad\Longleftrightarrow\quad
D < \min\{D^{\mathrm{high}}_{\pm}, D^\ast\}.
\]

These two together imply that the strict Pareto-improvement condition is:
\[
D < \min\{D^{\mathrm{high}}_0,\;D^{\mathrm{high}}_{\pm},\;D^\ast\},
\]
where (i) the popularity-driven equilibrium exists and is unique (since $D<D^\ast$ and $D^\ast > 0$), and
(ii) $\Delta U^0>0$ and $\Delta U^\pm>0$, so all agents are strictly better off relative to the
authentic benchmark.

\smallskip
\noindent\textbf{Existence of the Pareto-improving region.}
If $D^\ast>0$, $D^{\mathrm{high}}_0>0$, and $D^{\mathrm{high}}_{\pm}>0$, then
$
\min\{D^{\mathrm{high}}_0,\,D^{\mathrm{high}}_{\pm},\,D^\ast\}>0,
$
so the interval
$
\bigl(0,\min\{D^{\mathrm{high}}_0,\,D^{\mathrm{high}}_{\pm},\,D^\ast\}\bigr)
$
is nonempty. As shown in the proof of Proposition~\ref{cor:neutral_trap}, $D^{\mathrm{high}}_0<D^\ast$,
which implies that this interval simplifies to
\[
\bigl(0,\min\{D^{\mathrm{high}}_0,\,D^{\mathrm{high}}_{\pm}\}\bigr).
\]
Moreover, Proposition \ref{prop:everyone_worse_off} already shows since $G_0<\frac{n}{3}$ implies $D^{\mathrm{high}}_{\pm}>0$ holds, so that it suffices to ensure
that $D^{\mathrm{high}}_0>0$ for the existence of a strictly utility-improving $D$ region for everyone. Recall that $D_0^{\text{high}}
=
\frac{\omega^p (n-3G_0) - 2\omega^a G_0(aG_0+1)}
{(n-G_0)(aG_0+1)}.$ Then, given that the denominator of $D_0^{\text{high}}$is always positive, $D_0^{\text{high}}>0$ if and only if

\[
\omega^p> \overline{\omega}^p = \frac{2\omega^a G_0(aG_0+1)}{n-3G_0}.
\]

Hence, $\omega^p> \overline{\omega}^p$ guarantees both $D^\ast>0$ and $D^{\mathrm{high}}_0>0$. Therefore, for any
\[
D\in\bigl(0,\min\{D^{\mathrm{high}}_0,\,D^{\mathrm{high}}_{\pm},\,D^\ast\}\bigr),
\]
the popularity-driven equilibrium is unique and strictly utility-improving for everyone relative to
the authentic benchmark.

\end{proof}

\subsubsection*{Proof of Proposition \ref{prop:lowpol-D-regions}}

\begin{proof}
Fix a low-polarization event with $G_0>\frac{n}{3}$ and $G_-=G_+=\frac{n-G_0}{2}$.

By Proposition~\ref{prop1} (posting incentives), an opinionated agent prefers to post
neutral content $c_i=0$ rather than her authentic opinion $c_i=b_i\in\{\pm b\}$ if and only if

\[(\omega^p+\omega^a) a \frac{n-G_0}{2} < (\omega^p-D)aG_0\]

This inequality is equivalent to

\[D < D^{\ast\ast} = \frac{\omega^p (3 G_0 -n )- \omega^a(n-G_0) }{2 G_0 } \]

$D^{\ast\ast}>0$ holds if and only if $\omega^p > (\omega^p)^{\ast\ast} =\tfrac{\omega^a (n-G_0)}{3G_0-n}$. If $D > D^{**}(G_0)$,
all agents post authentically and the equilibrium coincides with the authentic benchmark.
Hence, whenever $D<D^{**}$, the popularity-driven equilibrium exists and is unique. 

Suppose $D < D^{\ast\ast}$ holds.

\paragraph{Utility comparison for opinionated agents and the threshold $D_{\pm}^{\text{low}}$.}

If $i \in \mathcal{G}_+$(symmetric for $i \in \mathcal{G}_-$) agent $i$ posts authentic $c_i = b_i$, the total popularity in her feed associated with each content type is
\[
S_i^{+b}=(aG_+ + 1)\cdot aG_+,\qquad
S_i^{0}=(aG_0)\cdot aG_0,\qquad
S_i^{-b}=(aG_-)\cdot aG_-.
\]

Then, consider the authentic benchmark outcome. Her utility gain under the authentic benchmark relative to the autarky benchmark can be written as:
\begin{align}
 \Delta U^{\mathrm{auth}}_\pm
&=\omega^p\Big(a\frac{n-G_0}{2}\Big)
+\omega^a\Big(a\frac{n-G_0}{2}+1\Big)\Big(a\frac{n-G_0}{2}\Big)
-D\Big(a^2G_0^2 + 2a^2\Big(\frac{n-G_0}{2}\Big)^2\Big).
\label{eq:auth-payoff-lowpol}
\end{align}

\smallskip
\noindent\emph{ Popularity-driven (neutral) posting.}
If $i \in \mathcal{G}_+$ instead posts $c_i=0\neq b_i$—and by symmetry all opinionated agents do so—the corresponding
popularity terms in her exposure set are
\[
S_i^{+b}=(aG_+)\cdot aG_+,\qquad
S_i^{0}=(aG_0+1)\cdot aG_0,\qquad
S_i^{-b}=(aG_-)\cdot aG_-.
\]
The resulting utility gain at equilibrium (relative to the autarky benchmark) is
\begin{align}
\Delta U^\mathrm{eq}_\pm
&=\omega^p(aG_0)
-D\Big((an+1)aG_0\Big).
\label{eq:dev-payoff-lowpol}
\end{align}

\paragraph{Comparison.}
Strategic neutral posting is utility-enhancing for opinionated agents if and only if
\begin{align*}
\Delta U^{\mathrm{eq}}_\pm - U^{\mathrm{auth}}_\pm
&=
\omega^p\!\left(aG_0 - a\frac{n-G_0}{2}\right)
-\omega^a\!\left(a\frac{n-G_0}{2}+1\right)\!\left(a\frac{n-G_0}{2}\right) \\
&\quad
-D\!\left((an+1)aG_0 - a^2 G_0^2 - 2a^2\left(\frac{n-G_0}{2}\right)^2\right).
\end{align*}

which can be rearranged as
\[
D
<
\frac{\omega^p a\big(G_0-\frac{n-G_0}{2}\big)
-\omega^a\big(a\frac{n-G_0}{2}+1\big)\big(a\frac{n-G_0}{2}\big)}
{(an+1)aG_0 - a^2 G_0^2 - 2a^2\left(\frac{n-G_0}{2}\right)^2}
\]

By expanding the terms and rewriting it:

\[
D
< D^{\mathrm{low}}_\pm =
\frac{\omega^p (3G_0-n)
-\omega^a\big(a\frac{n-G_0}{2}+1\big)(n-G_0)}
{2G_0 + a(n-G_0)(3G_0-n)}
\]

Recall that $$D^{**}(G_0)
= \frac{G_0(3\omega^p+\omega^a)-(\omega^p+\omega^a)n}{2G_0}=\frac{\omega^p (3G_0-n)-\omega^a(n-G_0)}{2G_0}$$. 
We first show that $
D^{\mathrm{low}}_\pm
<
D^{**}(G_0).
$

The numerator of $D^{\mathrm{low}}_\pm$ is strictly smaller than that of $D^{**}$, and since the denominator is strictly larger, we conclude that $D^{\mathrm{low}}_\pm < D^{**}(G_0)$.

Next, we find the $w_p$ threshold to have $D^{low}_{\pm}>0$.

$D^{low}_{\pm}>0$ if and only if

\[\omega^p > \tilde{\omega}^p =  \frac{\omega^a\big(a\frac{n-G_0}{2}+1\big)\big(n-G_0\big)}{\big(3G_0-n\big)}\]

For opinionated agents, $\Delta U^{\mathrm{eq},\pm}>\Delta U^{\mathrm{auth},\pm}$ holds if and only if
$D<D_{\pm}^{\mathrm{low}}$. Hence, a nonempty region in which opinionated agents are strictly better off
requires $D_{\pm}^{\mathrm{low}}>0$, which is equivalent to $\omega^p>\tilde{\omega}^p$.
By contrast, opinionated agents are strictly worse off whenever $D_{\pm}^{\mathrm{low}}<D<D^{**}$.
This region exists whenever the equilibrium is popularity-driven, where $D^{**}>0$ (or
$\omega^p>(\omega^p)^{**}$). 

\paragraph{Neutral agents are strictly better off under popularity equilibrium.}
Fix a neutral agent $j\in\mathcal G_0$ with $b_j=0$. Under the authentic benchmark,
$j$ posts $c_j=0$ and obtains popularity $R_j^{\mathrm{auth}}=aG_0$. Under the
popularity-driven equilibrium (where all opinionated agents post $0$), $j$  posts
$c_j=0$ and obtains the same popularity $R_j^{\mathrm{eq}}=aG_0$.

Under authentic posting by everyone, $j$ is exposed to $aG_0$ neutral posts from others
plus her own post, so aligned exposure equals $(aG_0+1)aG_0$. In addition, $j$ is exposed
to $aG_+=a\frac{n-G_0}{2}$ posts of each type $\pm b$, each at distance $|b|$, implying
total misaligned exposure $a^2(G_+^2+G_-^2)=a^2\frac{(n-G_0)^2}{2}$. Hence
\[
\Delta U^{\mathrm{auth}}_{0}
=
\omega^p aG_0
+\omega^a(aG_0+1)aG_0
-
D\,a^2\frac{(n-G_0)^2}{2}.
\]

Under the popularity-driven equilibrium, all posts in $j$'s feed are neutral. Therefore
misaligned exposure is zero, while aligned exposure remains $(aG_0+1)aG_0$, and
\[
\Delta U^{\mathrm{eq}}_{0}
=
\omega^p aG_0
+\omega^a(an+1)aG_0.
\]
Thus,
\[
\Delta U^{\mathrm{eq}}_{0}-\Delta U^{\mathrm{auth}}_{0}
=
D\,a^2\frac{(n-G_0)^2}{2}
>0,
\]
so neutral agents are always strictly better off under the popularity-driven equilibrium.

\end{proof}

\subsubsection*{Proof of Proposition \ref{prop:aggregate-welfare}}

\begin{proof}
Normalize autarky welfare to $\mathcal W^{\mathrm{autarky}}=\sum_{i\in\mathcal N}H_i=0$ where $H_i=\overline{H}=0$. Define
aggregate welfare by
\[
\mathcal W(\mathbf c,\mathbf R):=\sum_{i\in\mathcal N}(U_i).
\]

A neutral agent’s utility under authentic expression is
\[
U^{\mathrm{auth}}_0
=
\omega^p aG_0
+\omega^a(aG_0+1)aG_0
-
D\,a^2\frac{(n-G_0)^2}{2}.
\]

An opinionated agent’s utility (either $+b$ or $-b$) under authentic expression is
\[
U^{\mathrm{auth}}_\pm 
=
\omega^p\Big(a\frac{n-G_0}{2}\Big)
+\omega^a\Big(a\frac{n-G_0}{2}+1\Big)\Big(a\frac{n-G_0}{2}\Big)
-D\Big(a^2G_0^2 + 2a^2\Big(\frac{n-G_0}{2}\Big)^2\Big)
\]
Aggregate authentic welfare is therefore:

\[
\mathcal W^{\mathrm{auth}}
=
G_0\bigl(U^{\mathrm{auth}}_0\bigr)
+
(n-G_0)\bigl(U^{\mathrm{auth}}_\pm\bigr).
\]

Equivalently, written fully:

{\footnotesize

\begin{multline*}
\mathcal W^{\mathrm{auth}}
=
\omega^p a\!\left[
G_0^2
+
\frac{(n-G_0)^2}{2}
\right]
+\omega^a a\!\left[
G_0^2(aG_0+1)
+
\frac{(n-G_0)^2}{2}\!\left(
a\frac{n-G_0}{2}+1
\right)
\right]
\\
-
Da^2\!\left[
\frac{n(n-G_0)^2}{2}
+ 
\Big(G_0^2\Big)(n-G_0)
\right].
\end{multline*}
}

\paragraph{Popularity-driven equilibrium (high polarization). $D<D^\ast$}
Define aggregate equilibrium welfare (relative to $\overline H$) as
\[
\mathcal W^{\mathrm{eq}}
=
G_0\bigl(U^{\mathrm{eq}}_0-\overline H\bigr)
+
(n-G_0)\bigl(U^{\mathrm{eq}}_\pm-\overline H\bigr).
\]
In the popularity-driven equilibrium, neutral agents post $\pm b$ and all opinionated
posts receive popularity $A=a\frac{n-G_0}{2}$. 
\[
U^{\mathrm{eq}}_0
=
\omega^p a\frac{n-G_0}{2}
-
D(an+1)a\frac{n-G_0}{2},
\]
and
\[
U^{\mathrm{eq}}_\pm
=
\omega^p a\frac{n-G_0}{2}
+
\omega^a\Big(\frac{an}{2}+1\Big)a\frac{n-G_0}{2}
-
D(an)a\frac{n-G_0}{2}.
\]

Then,
\[
\mathcal W^{\mathrm{eq}}
=
n\omega^p\Big(a\frac{n-G_0}{2}\Big)
+
(n-G_0)\omega^a\Big(\frac{an}{2}+1\Big)\Big(a\frac{n-G_0}{2}\Big)
-
D\bigl(G_0(an+1)+an(n-G_0)\bigr)\Big(a\frac{n-G_0}{2}\Big).
\]

Let $\Delta \mathcal{W} = \mathcal{W}^{\mathrm{eq}} - \mathcal{W}^{\mathrm{auth}}$. We decompose the welfare expressions into three components based on the parameters $\omega^p$, $\omega^a$, and $D$.
Let $A = a\frac{n-G_0}{2}$.

\paragraph{1. The Popularity Component ($\omega^p$)}
The popularity term in the equilibrium welfare is:
\[
\mathcal{W}^{\mathrm{eq}}_p = n\omega^p A = \omega^p \frac{an(n-G_0)}{2}.
\]
The popularity term in the authentic benchmark is:
\[
\mathcal{W}^{\mathrm{auth}}_p = \omega^p a \left[ G_0^2 + \frac{(n-G_0)^2}{2} \right] 
= \frac{\omega^p a}{2} \left[ 2G_0^2 + n^2 - 2nG_0 + G_0^2 \right] 
= \frac{\omega^p a}{2} (3G_0^2 - 2nG_0 + n^2).
\]
Taking the difference $\Delta_p = \mathcal{W}^{\mathrm{eq}}_p - \mathcal{W}^{\mathrm{auth}}_p$:
\begin{align*}
\Delta_p &= \frac{\omega^p a}{2} \left[ (n^2 - nG_0) - (3G_0^2 - 2nG_0 + n^2) \right] \\
&= \frac{\omega^p a}{2} (nG_0 - 3G_0^2) \\
&= \frac{a G_0}{2} \omega^p (n - 3G_0).
\end{align*}

\paragraph{2. The Alignment Component ($\omega^a$)}
The equilibrium alignment term is:
\[
\mathcal{W}^{\mathrm{eq}}_a = (n-G_0)\omega^a \left(\frac{an}{2}+1\right) A 
= \omega^a \frac{a(n-G_0)^2}{2} \left(\frac{an}{2}+1\right).
\]
The authentic alignment term is:
\[
\mathcal{W}^{\mathrm{auth}}_a = \omega^a a \left[ G_0^2(aG_0+1) + \frac{(n-G_0)^2}{2}\left(A+1\right) \right].
\]
Subtracting the terms involving $(n-G_0)^2$ first:
\begin{align*}
\text{Diff}_{(n-G_0)} &= \omega^a \frac{a(n-G_0)^2}{2} \left[ \left(\frac{an}{2}+1\right) - \left(a\frac{n-G_0}{2}+1\right) \right] \\
&= \omega^a \frac{a(n-G_0)^2}{2} \left[ \frac{a}{2}(n - (n-G_0)) \right] \\
&= \omega^a \frac{a(n-G_0)^2}{2} \left( \frac{aG_0}{2} \right) = \frac{\omega^a a^2 G_0 (n-G_0)^2}{4}.
\end{align*}
Now subtracting the remaining authentic term ($G_0^2$ part):
\begin{align*}
\Delta_a &= \frac{\omega^a a^2 G_0 (n-G_0)^2}{4} - \omega^a a G_0^2(aG_0+1).
\end{align*}
Factor out $\frac{a G_0}{2} \cdot \frac{\omega^a}{2}$:
\begin{align*}
\Delta_a &= \frac{a G_0}{2} \frac{\omega^a}{2} \left[ a(n-G_0)^2 - 4G_0(aG_0+1) \right] \\
&= \frac{a G_0}{2} \frac{\omega^a}{2} \left[ a(n^2 - 2nG_0 + G_0^2) - 4aG_0^2 - 4G_0 \right] \\
&= \frac{a G_0}{2} \frac{\omega^a}{2} \left[ a(n^2 - 2nG_0 - 3G_0^2) - 4G_0 \right].
\end{align*}
Using the factorization $n^2 - 2nG_0 - 3G_0^2 = (n-3G_0)(n+G_0)$:
\[
\Delta_a = \frac{a G_0}{2} \frac{\omega^a}{2} \Big[ a(n-3G_0)(n+G_0) - 4G_0 \Big].
\]

\paragraph{3. The Misalignment Component ($D$)}
The equilibrium term coefficient for $-D$ is:
\begin{align*}
C^{\mathrm{eq}}_D &= A \left[ G_0(an+1) + an(n-G_0) \right] \\
&= \frac{a(n-G_0)}{2} \left[ anG_0 + G_0 + an^2 - anG_0 \right] \\
&= \frac{a(n-G_0)}{2} (an^2 + G_0).
\end{align*}
The authentic term coefficient for $-D$ is:
\begin{align*}
C^{\mathrm{auth}}_D &= a^2 \left[ \frac{n(n-G_0)^2}{2} + G_0^2(n-G_0) \right] \\
&= \frac{a^2(n-G_0)}{2} \left[ n(n-G_0) + 2G_0^2 \right] \\
&= \frac{a^2(n-G_0)}{2} (n^2 - nG_0 + 2G_0^2).
\end{align*}
The difference is $\Delta_D = -D (C^{\mathrm{eq}}_D - C^{\mathrm{auth}}_D)$:
\begin{align*}
C^{\mathrm{eq}}_D - C^{\mathrm{auth}}_D &= \frac{a(n-G_0)}{2} \left[ (an^2 + G_0) - a(n^2 - nG_0 + 2G_0^2) \right] \\
&= \frac{a(n-G_0)}{2} \left[ an^2 + G_0 - an^2 + anG_0 - 2aG_0^2 \right] \\
&= \frac{a(n-G_0)}{2} \left[ G_0 + anG_0 - 2aG_0^2 \right] \\
&= \frac{a(n-G_0)G_0}{2} \left[ 1 + a(n-2G_0) \right].
\end{align*}
Thus:
\[
\Delta_D = - \frac{a G_0}{2} D (n-G_0) \Big[ a(n-2G_0) + 1 \Big].
\]

Summing $\Delta_p$, $\Delta_a$, and $\Delta_D$ and factoring out $\frac{a G_0}{2}$:
\[
\Delta \mathcal{W} = \frac{a G_0}{2} \left[ 
    \omega^p(n-3G_0) 
    + \frac{\omega^a}{2}\Big( a(n-3G_0)(n+G_0) - 4G_0 \Big) 
    - D(n-G_0)\Big( a(n-2G_0) + 1 \Big) 
\right].
\]

Then by further algebra:
{\footnotesize
\[
\Delta \mathcal{W} = \mathcal{W}^{\mathrm{eq}} - \mathcal{W}^{\mathrm{auth}}
=
\frac{a G_0}{2} \left[ 
    \omega^p(n-3G_0) 
    + \frac{\omega^a}{2}\Big( a(n-3G_0)(n+G_0) - 4G_0 \Big) 
    - D(n-G_0)\Big( a(n-2G_0) + 1 \Big) 
\right]
\]}

\[
\Delta \mathcal W = \frac{aG_0}{2} \left[ \omega^p (n - 3G_0) 
+ \omega^a \left( \frac{a (n - G_0)^2}{2} - 2G_0(1 + aG_0) \right) 
+ D (n - G_0) \big( a(2G_0 - n) - 1 \big) \right].
\]

The equilibrium generates higher aggregate welfare than the authentic benchmark if and only if $D < \widehat{D}$, where:
\[
\widehat{D} = \frac{\omega^p(n-3G_0) + \frac{\omega^a}{2}\Big[ a(n-3G_0)(n+G_0) - 4G_0 \Big]}{(n-G_0)\Big[ a(n-2G_0) + 1 \Big]}.
\]

\[\omega^p(n-3G_0) + \frac{\omega^a}{2}\Big[ a(n-3G_0)(n+G_0) - 4G_0 \Big] > 0\]

Then, $\widehat{D}>0$ holds iff
\[
\widehat{\frac{\omega^p}{\omega^a}} > \frac{\Big[ a(n-3G_0)(n+G_0) - 4G_0 \Big]}{2(n-3G_0)}.\]

Recall that $D^*=\frac{\omega^p (n-3G_0) - 2\omega^a G_0}{n-G_0}$.

Then, $\widehat{D} < D^\ast$ holds if and only if: 

\[
\left(\frac{\omega^p}{\omega^a} \right)>  \left(\frac{\omega^p}{\omega^a} \right)^{\prime} = \frac{(n-3G_0)(n+G_0) + 4G_0(n-2G_0)}{2(n-3G_0)(n-2G_0)}.\]

We already know that $D^*$ is positive if and only if $\left(\frac{\omega^p}{\omega^a} \right)>\left(\frac{\omega^p}{\omega^a}\right)^*=\frac{2G_0}{n-3G_0}$.

Then, we have three cases: 

(a) for $\frac{\omega^p}{\omega^a}$ greater than both thresholds, $D^*$ is positive and $\widehat{D} < D^\ast$, implying there exists a non-empty region $(\widehat{D}, D^\ast)$, and for $D \in (\widehat{D}, D^\ast)$, the unique equilibrium is popularity-driven and welfare is lower than the authenticity benchmark welfare. 

(b) for $\left(\frac{\omega^p}{\omega^a}\right)^*<\frac{\omega^p}{\omega^a}<\left(\frac{\omega^p}{\omega^a}\right)^\prime$, $D^*$ is positive but $\widehat{D} > D^\ast$, implying whenever the unique equilibrium is popularity-driven ($D<D^*$), it is always welfare-improving, otherwise it is equivalent as we already know. 

(c) for $\left(\frac{\omega^p}{\omega^a}\right)^\prime<\frac{\omega^p}{\omega^a}<\left(\frac{\omega^p}{\omega^a}\right)^*$, $D^*$ would be negative and hence equilibrium would be always authentic expression, however, there exists no such region. \[
\left(\frac{\omega^{p}}{\omega^{a}}\right)^{*}=\frac{2G_{0}}{n-3G_{0}},\qquad
\left(\frac{\omega^{p}}{\omega^{a}}\right)'=\frac{(n-3G_{0})(n+G_{0})+4G_{0}(n-2G_{0})}{2(n-3G_{0})(n-2G_{0})},
\]
and under high polarization ($0<G_{0}<n/3$) one can simplify
\[
\left(\frac{\omega^{p}}{\omega^{a}}\right)' - \left(\frac{\omega^{p}}{\omega^{a}}\right)^{*}
= \frac{n+G_{0}}{2(n-2G_{0})} > 0.
\]
Thus $\left(\frac{\omega^{p}}{\omega^{a}}\right)' > \left(\frac{\omega^{p}}{\omega^{a}}\right)^{*}$ always holds.

This means for $\left(\frac{\omega^p}{\omega^a}\right)>\max \{\left(\frac{\omega^p}{\omega^a}\right)^*, \left(\frac{\omega^p}{\omega^a}\right)^\prime\}$, there exists a $D>0$ for which the unique equilibrium is popularity-driven and welfare is lower than the authenticity benchmark welfare. Otherwise, the equilibrium is weakly-welfare improving and strictly welfare-improving if part (b) above is satisfied.

\paragraph{Popularity-driven equilibrium (low polarization). $D<D^{\ast\ast}$}
In the low-polarization popularity-driven equilibrium, opinionated agents post neutral content,
while neutral agents post authentically. The per-agent equilibrium utilities are
\[
U^{\mathrm{eq}}_0
=
\omega^p aG_0
+\omega^a(an+1)aG_0
\]
and
\[
U^{\mathrm{eq}}_\pm
=
\omega^p aG_0
-
D(an+1)aG_0.
\]

\begin{multline*}
\mathcal W^{\mathrm{eq}}
=
G_0 U^{\mathrm{eq}}_0
+
(n-G_0)U^{\mathrm{eq}}_\pm \\
=
G_0\!\left[
\omega^p aG_0
+\omega^a(an+1)aG_0
\right]
+
(n-G_0)\!\left[
\omega^p aG_0
-
D(an+1)aG_0
\right].
\end{multline*}

\[
\mathcal W^{\mathrm{eq}}
=
\omega^p a\,nG_0
+\omega^a a(an+1)G_0^2
-
D(an+1)aG_0(n-G_0).
\]

\begin{multline*}
\mathcal W^{\mathrm{auth}}
=
\omega^p a\!\left[
G_0^2
+
\frac{(n-G_0)^2}{2}
\right]
+\omega^a a\!\left[
G_0^2(aG_0+1)
+
\frac{(n-G_0)^2}{2}\!\left(
a\frac{n-G_0}{2}+1
\right)
\right]
\\
-
Da^2\!\left[
\frac{n(n-G_0)^2}{2}
+ 
\Big(G_0^2\Big)(n-G_0)
\right].
\end{multline*}

Let $\Delta \mathcal{W} = \mathcal{W}^{\mathrm{eq}} - \mathcal{W}^{\mathrm{auth}}$. We calculate the difference term by term.\paragraph{1. The Popularity Component ($\omega^p$)}$$\mathcal{W}^{\mathrm{eq}}_p = \omega^p a n G_0.$$$$\mathcal{W}^{\mathrm{auth}}_p = \omega^p a \left[ G_0^2 + \frac{(n-G_0)^2}{2} \right] 
= \frac{\omega^p a}{2} \left[ 2G_0^2 + n^2 - 2nG_0 + G_0^2 \right] 
= \frac{\omega^p a}{2} (3G_0^2 - 2nG_0 + n^2).$$

By rewriting the terms in brackets:
 
\begin{align*}\Delta_p = \mathcal{W}^{\mathrm{eq}}_p - \mathcal{W}^{\mathrm{auth}}_p= \frac{\omega^p a}{2} (n-G_0)(3G_0-n).\end{align*}

\paragraph{2. The Alignment Component ($\omega^a$)}$$\mathcal{W}^{\mathrm{eq}}_a = \omega^a a (an+1) G_0^2.$$$$\mathcal{W}^{\mathrm{auth}}_a = \omega^a a \left[ G_0^2(aG_0+1) + \frac{(n-G_0)^2}{2}\left(\frac{a(n-G_0)}{2}+1\right) \right].$$

By rewriting the terms in brackets:

$$\Delta_a = \frac{\omega^a a (n-G_0)}{4} \Big[ a(3G_0-n)(n+G_0) - 2(n-G_0) \Big].$$\paragraph{3. The Misalignment Component ($D$)}$$\mathcal{W}^{\mathrm{eq}}_D = -D a G_0 (n-G_0) (an+1).$$$$\mathcal{W}^{\mathrm{auth}}_D = -D a^2 \left[ \frac{n(n-G_0)^2}{2} + G_0^2(n-G_0) \right] 
= -D a^2 (n-G_0) \left[ \frac{n(n-G_0)}{2} + G_0^2 \right].$$

By rewriting the terms in brackets:

$$\Delta_D = - \frac{D a (n-G_0)}{2} \Big[ 2G_0 + a(n-G_0)(2G_0-n) \Big].$$

By summing up all: 

{\footnotesize
$$\Delta \mathcal{W} = \frac{a(n-G_0)}{2} \left[ 
    \omega^p (3G_0-n) 
    + \frac{\omega^a}{2} \Big( a(3G_0-n)(n+G_0) - 2(n-G_0) \Big) 
    - D \Big( 2G_0 + a(n-G_0)(2G_0-n) \Big) 
\right].$$
}

The equilibrium generates higher aggregate welfare than the authentic benchmark if and only if $D < \widehat{D}^{\mathrm{low}}$, where:
\[
\widehat{D}^{\mathrm{low}} = \frac{\omega^p(3G_0-n) + \frac{\omega^a}{2}\Big[ a(3G_0-n)(n+G_0) - 2(n-G_0) \Big]}{2G_0 + a(n-G_0)(2G_0-n)}.
\]

The condition $\widehat{D}^{\mathrm{low}} > 0$ holds if the numerator is positive. Since $3G_0-n > 0$, this is always true for sufficiently large $\omega^p$. Specifically, it holds if:
\[
\frac{\omega^p}{\omega^a} > \frac{2(n-G_0) - a(3G_0-n)(n+G_0)}{2(3G_0-n)}.
\]

Recall the existence threshold for the popularity-driven equilibrium is $D^{**} = \frac{\omega^p(3G_0-n) - \omega^a(n-G_0)}{2G_0}$.
We compare $\widehat{D}^{\mathrm{low}}$ and $D^{**}$.
$\widehat{D}^{\mathrm{low}} < D^{**}$ holds if and only if:
\[
\frac{\omega^p(3G_0-n) + \frac{\omega^a}{2}\Big[ a(3G_0-n)(n+G_0) - 2(n-G_0) \Big]}{2G_0 + a(n-G_0)(2G_0-n)}
<
\frac{\omega^p(3G_0-n) - \omega^a(n-G_0)}{2G_0}.
\]
% Let $X = 3G_0-n$, $Y = n-G_0$, and $Z = 2G_0-n$. The inequality is:
% \[
% \frac{\omega^p X + \frac{\omega^a}{2}[ a X (n+G_0) - 2Y ]}{2G_0 + a Y Z} < \frac{\omega^p X - \omega^a Y}{2G_0}.
% \]
% Cross-multiplying (assuming denominators are positive, i.e., $G_0 > n/2$ for the $Z$ term assurance, or simply assuming the term is positive as discussed):
% \[
% 2G_0 \left( \omega^p X + \frac{\omega^a}{2}[ a X (n+G_0) - 2Y ] \right) < (\omega^p X - \omega^a Y)(2G_0 + a Y Z).
% \]
% Expanding and collecting $\omega^p$ terms on the left and $\omega^a$ terms on the right:
% \[
% \omega^p X [ 2G_0 - (2G_0 + a Y Z) ] < \omega^a \left[ -Y(2G_0 + a Y Z) - G_0(a X (n+G_0) - 2Y) \right].
% \]
% \[
% \omega^p X [ - a Y Z ] < \omega^a \left[ -2G_0 Y - a Y^2 Z - a G_0 X (n+G_0) + 2G_0 Y \right].
% \]
% The $-2G_0 Y$ and $+2G_0 Y$ terms cancel on the RHS:
% \[
% - \omega^p a X Y Z < - \omega^a a \left[ Y^2 Z + G_0 X (n+G_0) \right].
% \]
% Multiplying by $-1$ reverses the inequality:
% \[
% \omega^p a X Y Z > \omega^a a \left[ Y^2 Z + G_0 X (n+G_0) \right].
% \]
% Dividing by $a X Y Z$ (assuming $Z = 2G_0-n > 0$):
% \[
% \frac{\omega^p}{\omega^a} > \frac{Y^2 Z + G_0 X (n+G_0)}{X Y Z} = \frac{Y}{X} + \frac{G_0(n+G_0)}{Y(2G_0-n)}.
% \]

By further algebra, we have this threshold level as:

\[
\left(\frac{\omega^p}{\omega^a}\right)^{\prime\prime} := \frac{n-G_0}{3G_0-n} + \frac{G_0(n+G_0)}{(n-G_0)(2G_0-n)}.
\]

Thus, we have three cases for the Low Polarization regime:

\begin{itemize}
\item for $\frac{\omega^p}{\omega^a}>\max\left\{\left(\frac{\omega^p}{\omega^a}\right)^{\ast\ast}, \left(\frac{\omega^p}{\omega^a}\right)^{\prime \prime}\right\}$ greater than both thresholds; $D^{\ast\ast}$ is positive and $D^{\prime\prime} < D^{\ast\ast}$, the region $(D^{\prime\prime}, D^\ast)$ is non-empty. Then, for $\frac{\omega^p}{\omega^a}>\max\left\{\left(\frac{\omega^p}{\omega^a}\right)^{\ast\ast}, \left(\frac{\omega^p}{\omega^a}\right)^{\prime \prime}\right\}$ and for $D>0$ satisfying $D \in (D^{\prime\prime}, D^\ast)$, the unique equilibrium is popularity-driven and welfare is lower than the authenticity benchmark welfare. 

\item for $\left(\frac{\omega^p}{\omega^a}\right)^{\ast\ast}<\frac{\omega^p}{\omega^a}<\left(\frac{\omega^p}{\omega^a}\right)^{\prime\prime}$; $D^{\ast\ast}$ is positive but $D^{\prime\prime} > D^{\ast\ast}$, implying whenever the unique equilibrium is popularity-driven $0<D<D^{\ast\ast}$, the equilibrium is welfare-improving; otherwise for $D \geq D^{\ast\ast}$, the equilibrium outcome is authentic expression. 

\item for $\left(\frac{\omega^p}{\omega^a}\right)^{\prime\prime}<\frac{\omega^p}{\omega^a}<\left(\frac{\omega^p}{\omega^a}\right)^{\ast\ast}$, $D^*$ would be negative, and, thus the equilibrium would be authentic expression. However, $(\omega^{p}/\omega^{a})^{\prime\prime}$ equals $(\omega^{p}/\omega^{a})^{**}$ plus a strictly positive term whenever $G_{0}>n/2$, so the inequality always holds for $G_{0}>n/2$, eliminating this region.  
\end{itemize}

\end{proof}

\subsubsection*{Proof of Proposition~\ref{proposition:algorithm}}\begin{proof}

We compare outcomes under RA and PVM.
\noindent\textit{ Likes under RA and PVM.}

\smallskip
\noindent\emph{RA:} Under the representative algorithm with visibility cap $k$, a post $c\in \mathcal{O}$
is shown to $k$ users with representative composition. Hence the total number of likes is
$
R^{RA}(c)=k\frac{G_c}{n}.
$

\smallskip
\noindent\emph{PVM:} Under post--viewer--match, a post $c_i\in \mathcal{O}$ is shown only to users whose
authentic opinion is same as the view in $c_i$, up to the cap $k$. Hence
$
R^{PVM}(c_i)=\min\{k,G_{c_i}\}.
$

\medskip
\textit{High-polarization event under RA.}
Suppose $0<G_0<\frac{n}{3}$. Consider a neutral agent ($b_i=0$).
Posting $c_i = 0$ yields:
\[
U^{RA}_i=\overline{H}+F_i+(\omega^p+\omega^a)\,k\frac{G_0}{n},
\]
whereas posting an opinionated content $c_i \in\{+b,-b\}$ yields:
\[
U^{RA}_i= \overline{H} +F_i+(\omega^p-\omega^d|b|)\,k\frac{n-G_0}{2n}.
\]
Therefore, a neutral agent posts opinionated under RA iff 

\[
(\omega^p-\omega^d|b|)\frac{n-G_0}{2}>(\omega^p+\omega^a)G_0.
\]

Thus, neutral agents post opinionated iff
\begin{equation*}\label{eq:Gstar}
G_0<G^\ast = \frac{(\omega^p-\omega^d|b|)n}{3\omega^p+2\omega^a-\omega^d|b|}.
\end{equation*}

\noindent\textit{High-polarization event  under PVM.}
Under post--viewer match (PVM), a post is shown only to aligned viewers, up to
the cap $k$, so $R^{PVM}(c)=\min\{k,G_c\}$. Therefore, for a zero-type agent,
posting $c_i \in\{+b,-b\}$ is profitable if and only if
\begin{equation*}
\overline{H} + F_i +(\omega^p+\omega^a)\min\{k,G_0\}
<
\overline{H} + F_i +(\omega^p-\omega^d|b|)\min\!\left\{k,\frac{n-G_0}{2}\right\}.
\label{eq:PVM-dev}
\end{equation*}
Let $g:=\frac{n-G_0}{2}$.
\noindent\underline{Case 1: $k\le G_0$.}
Then $\min\{k,G_0\}=k$. Since $G_0 < n$ implies $g\ge 0$ and, in particular,
$k\le G_0$ entails $\min\{k,g\}=k$ whenever $k \le g$. In that subcase,
the comparison becomes $(\omega^p+\omega^a)k<(\omega^p-\omega^d|b|)k$, which
is impossible because $\omega^a>0$. Hence, PVM induces no strategic posting whenever
$k\le G_0$.
\noindent\underline{Case 2: $G_0<k\le g$.}
Then $\min\{k,G_0\}=G_0$ and $\min\{k,g\}=k$, so the strategic posting condition reduces to
\[
(\omega^p+\omega^a)G_0<(\omega^p-\omega^d|b|)k,
\]
which is equivalent to

\begin{equation*}
k>
G_0\cdot\frac{\omega^p+\omega^a}{\omega^p-\omega^d|b|}.
\label{eq:kstar_high}
\end{equation*}

\noindent\underline{Case 3: $k>g$.}
Then $\min\{k,G_0\}=G_0$ and $\min\{k,g\}=g$, the strategic posting condition is exactly the same as under RA. Thus, in this case PVM induces
strategic posting if and only if $G_0<G^\ast$.

Thus, whenever popularity-driven posting occurs under RA, its occurrence under
PVM depends on $k$. If
$k< G_0\cdot\frac{\omega^p+\omega^a}{\omega^p-\omega^d|b|},$
then no such strategic posting arises under PVM. Hence,
$
C_0^{RA}\le C_0^{PVM},
$
and with strict inequality $
C_0^{RA} < C_0^{PVM},
$ if and only if
\[
G_0<G^\ast
\quad\text{and}\quad
k<
G_0\cdot\frac{\omega^p+\omega^a}{\omega^p-\omega^d|b|}.
\]

\noindent\textit{ Low-polarization event under RA.}
Suppose $\frac{n}{3}<G_0<n$. Consider an opinionated agent ($b_i\in\{\mathcal{G}_-,\mathcal{G}_+\}$) post authentic:
\[
U^{RA}(b_i)=\overline{H}+F_i+(\omega^p+\omega^a)\,k\frac{G_{b_i}}{n}
=H+F_i+(\omega^p+\omega^a)\,k\frac{n-G_0}{2n}.
\]
If opinionated agents create a popularity driven post $c_i = 0$
\[
U^{RA}(0)=\overline{H}+F_i+(\omega^p-\omega^d|b|)\,k\frac{G_0}{n}.
\]
Thus, opinionated agents post $b_i=0$ under RA iff
\[
(\omega^p-\omega^d|b|)\,k\frac{G_0}{n}>(\omega^p+\omega^a)\,k\frac{n-G_0}{2n},
\]

which is equivalent to 
\begin{equation*}\label{eq:Gstarstar}
G_0>G^{\ast\ast}\equiv \frac{(\omega^p+\omega^a)n}{3\omega^p-2\omega^d|b|+\omega^a}.
\end{equation*}

\noindent\textit{ Low-polarization event under PVM.} Under PVM the utility of opinionated agents are as follows: 
\[
U^{PVM}(b_i)=\overline{H}+F_i+(\omega^p+\omega^a)\min\left\{k,\frac{n-G_0}{2}\right\},
\]
and
\[
U^{PVM}(0)=\overline{H}+F_i+(\omega^p-\omega^d|b|)\min\{k,G_0\}.
\]

\noindent\underline{Case 1: $k\le g$.}
Then $\min\{k,g\}=k$ and (since $g\le G_0$ in this regime) also $\min\{k,G_0\}=k$.
In that subcase, the comparison becomes
$
(\omega^p+\omega^a)k<(\omega^p-\omega^d|b|)k,
$
 Hence, there is no strategic posting whenever $k<G_0$.
\noindent\underline{Case 2: $g<k\le G_0$.}
Then $\min\{k,g\}=g$ and $\min\{k,G_0\}=k$, and hence, strategic posting occurs iff
\[
(\omega^p+\omega^a)g<(\omega^p-\omega^d|b|)k,
\]
equivalently
\[
k>\frac{g(\omega^p+\omega^a)}{\omega^p-\omega^d|b|}=
\frac{n-G_0}{2}\cdot\frac{\omega^p+\omega^a}{\omega^p-\omega^d|b|}.
\]

\noindent\underline{Case 3: $k>G_0$.}
Then $\min\{k,G_0\}=G_0$ and, since $k>G_0\ge g$, we have $\min\{k,g\}=g$.
\[
(\omega^p+\omega^a)g<(\omega^p-\omega^d|b|)G_0,
\]
which is exactly the same as under RA strategic posting condition $G_0>G^{\ast\ast}$.

Thus, under PVM opinionated agents do not post neutral whenever
\begin{equation*}\label{eq:k_low}
(\omega^p-\omega^d|b|)k \le (\omega^p+\omega^a)\frac{n-G_0}{2}
\quad \Longleftrightarrow \quad
k \le \frac{(n-G_0)}{2} \cdot \frac{(\omega^p+\omega^a)}{(\omega^p-\omega^d|b|)}.
\end{equation*}

\end{proof}

\subsubsection*{A Numerical Example on Popularity Trap}

\label{sec:appendix_example}

In the numerical example provided in Table \ref{tab:three_opinions_combined}, we isolate this outcome by selecting parameters where the misalignment cost $D$ falls into a critical region. Specifically, the cost is low enough ($D < D^*$) to induce neutral agents to post  popularity- driven, yet high enough ($D > D_0^{\text{high}}$) that the resulting polarized environment lowers their utility. Simultaneously, the cost exceeds the threshold for opinionated agents ($D > D_\pm^{\text{high}}$), meaning the influx of aligned content fails to compensate them for the loss of neutral content.

\begin{table}[h]
\centering
\renewcommand{\arraystretch}{1} % Increases vertical row margins
\setlength{\tabcolsep}{2pt}       % Increases horizontal column margins
\caption{An Example of Utility Loss for Everyone in a High-Polarization Event.}
\label{tab:three_opinions_combined}
\begin{center}
\small
\textbf{Parameters:} $n=100$, $G_0=10$, $G_\pm=45$, $\omega^p=3$, $\omega^a=1$, $\omega^d=1$, $|b|=1.5$ ($D=1.5$).
\end{center}
\vspace{0.1cm}

\begin{tabular}{cccccccc}
\hline\hline
\multicolumn{8}{c}{\textbf{Utility Comparison}} \\
\hline
\textbf{Density} & \multicolumn{2}{c}{\textbf{Autarky}} & \multicolumn{2}{c}{\textbf{Authentic Benchmark}} & \multicolumn{2}{c}{\textbf{Equilibrium}} \\
($a$) & $\Delta U_\pm$ & $\Delta U_0$ & $\Delta U_\pm^{\mathrm{Auth}}$ & $\Delta U_0^{\mathrm{Auth}}$ & $\Delta U_\pm^{\mathrm{Eq}}$ & $\Delta U_0^{\mathrm{Eq}}$ & \\
\hline
$0.1$ & $0$ & $0$ & $-24.00$ & $-55.75$ & \cellcolor{gray!25}$-94.50$ & \cellcolor{gray!25}$-60.75$  \\
$0.4$ & $0$ & $0$ & $-600.00$ & $-940.00$ & \cellcolor{gray!25}$-1728.00$ & \cellcolor{gray!25}$-1053.00$ \\
\hline\hline
\end{tabular}

\vspace{0.2cm}
\footnotesize
\textbf{Extended Notes:}
\begin{itemize}
    \item \textbf{Threshold ($D^*$) for the popularity-driven equilibrium:} The  threshold is $D^* \approx 2.11$ (invariant to $a$). Since $D=1.5 < 2.11$, the equilibrium is popularity-driven for both cases $a=0.1$ and $a=0.4$.
    \item \textbf{Threshold Analysis ($a=0.1$):} The welfare loss thresholds are $D_0^{\mathrm{high}} \approx 0.94$ and $D_\pm^{\mathrm{high}} = 0.90$. The condition $\max\{0.94, 0.90\} < 1.5 < 2.11$ holds.
    \item \textbf{Threshold Analysis ($a=0.4$):} The welfare loss thresholds are $D_0^{\mathrm{high}} \approx 0.24$ and $D_\pm^{\mathrm{high}} = 0.60$. The condition $\max\{0.24, 0.60\} < 1.5 < 2.11$ holds.
    \item \textbf{Effect of Density:} Increasing connectivity relaxes the lower bound for welfare loss (from $0.94$ down to $0.60$), making the trap region larger, while simultaneously magnifying the magnitude of the utility loss (e.g., an opinionated agent's utility loss deepens from $-94.5$ to $-1728$).
\end{itemize}
\end{table}

\end{document}